\documentclass[12pt]{article}
\pdfoutput=1
\usepackage{putex}
\usepackage{feyn}
\usepackage[vcentermath]{youngtab}
\usepackage{subfig}

\usepackage{graphicx}
\usepackage{epstopdf}
\usepackage{enumerate}
\usepackage{cite}
\usepackage{tensor}
\usepackage{slashed}
\usepackage{amsmath}
\usepackage{amssymb}
\usepackage{mathrsfs}
\usepackage{lgrind}

\usepackage{feynmp-auto}

\usepackage{hyperref}

\numberwithin{equation}{section}

\newcommand{\OO}{\mathbb{O}}

\newcommand {\be} {\begin {equation}}
\newcommand {\ee} {\end {equation}}

\newcommand {\bes} {\begin {equation*}}
\newcommand {\ees} {\end {equation*}}

\newcommand{\Par}{\partial_{\mu}}

\newcommand{\eps}{\epsilon}

\def\CH{{\cal H}}

\newcommand{\beq}{\begin{equation}}
\newcommand{\eeq}{\end{equation}}

\def\be{ \begin{equation} }
\def\ee{ \end{equation} }

\begin{document}

\preprint{PUPT-2463}

\institution{PU}{Department of Physics, Princeton University, Princeton, NJ 08544}
\institution{PCTS}{Princeton Center for Theoretical Science, Princeton University, Princeton, NJ 08544}

\title{
Critical $O(N)$ Models in $6-\epsilon$ Dimensions
}

\authors{Lin Fei,\worksat{\PU} Simone Giombi\worksat{\PU} and Igor R.~Klebanov\worksat{\PU,\PCTS}
}

\abstract{We revisit the classic $O(N)$ symmetric scalar field theories in $d$ dimensions with interaction $(\phi^i \phi^i)^2$.
For $2<d<4$ these theories flow to the Wilson-Fisher fixed points for any $N$.
A standard large $N$ Hubbard-Stratonovich approach also indicates that, for $4<d<6$, these theories
possess unitary UV fixed points. We propose their alternate description in terms of a theory of $N+1$ massless scalars with the cubic interactions
$\sigma \phi^i \phi^i$ and $\sigma^3$. Our one-loop calculation in $6-\epsilon$ dimensions shows that this theory has
an IR stable fixed point at real values of the coupling constants for $N>1038$. We show that the $1/N$ expansions of various operator scaling dimensions match the known results for the critical $O(N)$ theory continued to $d=6-\epsilon$. These results suggest that, for sufficiently large $N$, there are 5-dimensional unitary $O(N)$ symmetric interacting CFT's; they
should be dual to the Vasiliev higher-spin theory in AdS$_6$ with alternate boundary conditions for the bulk scalar. Using these CFT's we provide a new test of the 5-dimensional $F$-theorem, and also find a new counterexample for the $C_T$ theorem.}

\date{}
\maketitle

\tableofcontents

\section{Introduction and Summary}

Among the many physical applications of Quantum Field Theory (QFT), an important role is played by its description of the second order phase transitions. These transitions are ubiquitous in statistical systems, such as the three-dimensional Ising model whose second order transition describes, for example, the critical point in the water-vapor phase diagram. The vicinity of the critical point may be described by the three-dimensional Euclidean QFT of a real scalar field, $\phi$, with a
$\lambda \phi^4$ interaction. Such a QFT is weakly coupled at short distances (in the UV), where the scaling dimension of the $\phi^4$ operator is $2$, but it becomes strongly coupled at long distances (in the IR). It is the long distance regime that is needed for describing the critical behavior, but perturbation theory in $\lambda$ cannot be used there.

Luckily, theorists have invented ingenious expansion schemes that have led to good approximations for the IR scaling dimensions of composite operators.
One of them is dimensional continuation: instead of working directly in $d=3$, it is fruitful to study the physics as a function of the dimension $d$. In the $\phi^4$ theory there is evidence that the IR critical behavior occurs for $2 < d < 4$, and significant simplification occurs for $d=4-\epsilon$ where $\epsilon \ll 1$. Then the IR stable fixed point of the Renormalization Group occurs for $\lambda$ of order $\epsilon$, so that a formal
Wilson-Fisher expansion in $\epsilon$ may be developed \cite{Wilson:1971dc}. The coefficients of the first few terms fall off rapidly, so that setting $\epsilon=1$
provides a rather precise approximation that is in good agreement with the experimental and numerical results \cite{Wilson:1973jj}.

Another important idea has been the large $N$ expansion in the $O(N)$ symmetric QFT of $N$ real scalar fields $\phi^i$, $i=1, \ldots, N$,
with interaction ${\lambda\over 4} (\phi^i \phi^i)^2$. For small values of $N$ there are physical systems whose critical behavior is described by this $d=3$ QFT.
Furthermore, using a generalized Hubbard-Stratonovich transformation with an auxiliary field $\sigma$, it is possible to develop expansion in powers of $1/N$
(for a comprehensive review, see \cite{Moshe:2003xn}). The large $N$ expansion may be developed for a range of $d$ \cite{Vasiliev:1981yc,Vasiliev:1981dg,Vasiliev:1982dc,Lang:1990ni, Lang:1991kp, Lang:1992pp, Lang:1992zw, Petkou:1994ad,Petkou:1995vu} and compared with the regimes where other perturbative
expansions are available. In particular, for $d=4-\epsilon$, the Wilson-Fisher $\epsilon$-expansion may be developed
for any $N$ \cite{Wilson:1973jj}, and the results have to match with the large $N$ techniques. Also, for $d=2+\epsilon$ the large $N$ results match with the perturbative UV fixed point of
the $O(N)$ Non-linear Sigma Model (NL$\sigma$M). Using the first few terms in the large $N$ expansion directly in $d=3$ provides another approach to estimating the scaling dimensions for low values of $N$. Thus, a combination of the large $N$ and $\epsilon$ expansions provides good approximations for the critical behavior in the entire range $2 < d < 4$. One should note, however, that both expansions are not convergent but rather provide asymptotic series. There are continued efforts towards obtaining a more rigorous approach to the $O(N)$ symmetric CFT's using conformal bootstrap ideas \cite{Polyakov:1974gs,Ferrara:1973yt,Rattazzi:2008pe,Rychkov:2011et}, and recently it has led to more precise numerical calculations of the operator scaling dimensions in three-dimensional CFT's \cite{ElShowk:2012ht,Kos:2013tga}. The bootstrap approach may, in fact, be applied in the entire range $2 < d < 4$ \cite{El-Showk:2013nia}.

In this paper we will discuss extensions of these results to interacting $O(N)$ models in $d>4$. At first glance, such extensions seem impossible:
the $(\phi^i \phi^i)^2$ interaction is irrelevant at the Gaussian fixed point, since it has scaling dimension $2d-4$. Therefore, the long-distance behavior of this
QFT is described by the free field theory. However, at least for large $N$, the theory possesses a UV stable fixed point whose existence may be demonstrated using
the Hubbard-Stratonovich transformation. At the interacting fixed point, the scaling dimension of the operator $\phi^i \phi^i$ is $2+{\cal O} (1/N)$ for any $d$.
For $d>6$, this dimension is below the unitarity bound $d/2-1$. It follows that the interesting range, where the UV fixed point may be unitary, is
\cite{Parisi:1975im,parisi1977non,Bekaert:2011cu,Bekaert:2012ux}
\begin{equation}
\label{intrange}
4 < d < 6 \ .
\end{equation}
We will examine the structure of the $O(N)$ symmetric scalar field theory in this range from various points of view.

We first note that the coefficients in the $1/N$ expansions derived for various quantities in \cite{Vasiliev:1981yc,Vasiliev:1981dg,Vasiliev:1982dc,Lang:1990ni, Lang:1991kp, Lang:1992pp, Lang:1992zw, Petkou:1994ad,Petkou:1995vu} may be
continued to the range (\ref{intrange}) without any obvious difficulty. Thus, the UV fixed points make sense at least in the $1/N$ expansion. In Section
\ref{largeNreview} we present a review and discussion of the $1/N$ expansion, with emphasis on the range $4<d<6$.
However, one should be concerned about the stability of the fixed points in this range of dimensions at finite $N$. In $d=4+\epsilon$, where the UV fixed point is weakly coupled, it occurs for the negative
quartic coupling $\lambda_*=- \frac {8 \pi^2} {N+8}\epsilon + O(\epsilon^2)$ \cite{Maldacena:2012sf,Giombi:2014iua}. Thus, it seems that the UV fixed point theory
is unlikely to be completely stable, although it may be metastable.
In order to gain a better understanding of the fixed point theory, it would be helpful to describe it via RG flow
from another theory. In such a ``UV complete" description, the $O(N)$ symmetric theory we are after should appear as the conventional IR stable fixed point.

Our main result is to demonstrate that such a UV completion indeed exists: it is the $O(N)$ symmetric theory of $N+1$ scalar fields with the Lagrangian
\begin{equation}
{\cal L} = \frac{1}{2}(\Par \phi^i)^2+\frac{1}{2}(\Par \sigma)^2+\frac{g_1}{2}\sigma \phi^i \phi^i + \frac{g_2}{6}\sigma^3\ .
\label{cubic6d}
\end{equation}
The cubic interaction terms are relevant for $d<6$, so that the theory flows from the Gaussian fixed point to an interacting IR fixed point. The latter is
expected to be weakly coupled for $d=6-\epsilon$. The idea to study a cubic scalar theory in $d=6-\epsilon$ is not new. Michael Fisher has explored such an $\epsilon$-expansion
in the theory of a single scalar field as a possible description of the Yang-Lee edge singularity in the Ising model \cite{Fisher:1978pf}. In that case,
which corresponds to the $N=0$ version of (\ref{cubic6d}), the IR fixed point is at an imaginary value of $g_2$.\footnote{Renormalization group calculations for similar cubic theories in $d=6-\epsilon$ were carried out in \cite{deAlcantaraBonfim:1980pe, deAlcantaraBonfim:1981sy,Barbosa:1986kv}.} This is related to the lack of unitarity of the
fixed point theory. Using the one-loop beta functions for $g_1$ and $g_2$, we will show in Section \ref{IRfixed} that for large $N$ the IR stable fixed point instead occurs for
{\it real} values of the couplings, thus removing conflict with unitarity. Remarkably, such unitary fixed points exist only for $N> 1038$.
As we show in Section \ref{finiteN}, for $N\leq 1038$ the IR stable fixed point becomes complex, and the theory is no longer unitary. Thus, the breakdown of the large $N$ expansion in $d=6-\epsilon$ occurs at a very large
value, $N_{\rm crit}=1038$. We will provide evidence, however, that for the physically interesting dimension $d=5$, $N_{\rm crit}$ is much smaller.

Besides its intrinsic interest, the $O(N)$ invariant scalar CFT in $d=5$ has interesting applications to higher spin AdS/CFT dualities.
There exists a class of Vasiliev theories in AdS$_{d+1}$ \cite{Fradkin:1987ks,Vasiliev:1990en,Vasiliev:1992av,Vasiliev:1995dn,Vasiliev:1999ba, Vasiliev:2003ev} that is naturally conjectured to be dual to the $O(N)$ singlet sector of the $d$-dimensional CFT of $N$ free
scalars \cite{Klebanov:2002ja,Giombi:2014iua}. In order to extend the duality to
interacting CFT's, one adds the $O(N)$ invariant term ${\lambda\over 4} (\phi^i \phi^i)^2$. For $d=3$ this leads to the well-known Wilson-Fisher fixed points \cite{Wilson:1971dc,Wilson:1973jj}.
In the dual description of these large $N$ interacting theories, it is necessary to change the $r^{-\Delta}$ boundary conditions on the scalar field in AdS$_4$ from the $\Delta_-=1$ to $\Delta_+=2+ {\cal O}(1/N)$ \cite{Klebanov:2002ja}. The situation is very similar for the $d=5$ case, which should be dual to the Vasiliev theory in AdS$_6$ \cite{Giombi:2014iua}. One can adopt the
$\Delta_+=3$ boundary conditions on the bulk scalar, which are necessary for the duality to the free $O(N)$ theory. Alternatively, the $\Delta_-=2+ {\cal O}(1/N)$
are allowed as well \cite{Breitenlohner:1982jf,Klebanov:1999tb}. This suggests that the dual interacting $O(N)$ CFT should exist in $d=5$, at least for large $N$ \cite{Maldacena:2012sf,Giombi:2014iua}. Our RG calculations lend further support
to the existence of this interacting $d=5$ CFT.

In Sections \ref{IRfixed} and \ref{opmix}, using one-loop calculations for the theory (\ref{cubic6d}) in $d=6-\epsilon$, we find some IR operator dimensions to order $\epsilon$, while keeping track of the dependence on $1/N$ to any desired order.
We will then match our results with the $1/N$ expansions derived for the $(\phi^i \phi^i)^2$ theory in \cite{Vasiliev:1981yc,Vasiliev:1981dg,Vasiliev:1982dc,Lang:1990ni, Lang:1991kp, Lang:1992pp, Lang:1992zw, Petkou:1994ad,Petkou:1995vu}, evaluating
them in $d=6-\epsilon$. The perfect match of the coefficients in these two $1/N$ expansions provides convincing evidence that the IR fixed point of the
cubic $O(N)$ theory (\ref{cubic6d}) indeed describes the same physics as the UV fixed point of the $(\phi^i \phi^i)^2$ theory. Our results thus provide evidence that,
at least for large $N$, the interacting unitary $O(N)$ symmetric scalar CFT's exist not only for $2<d<4$, but also for $4<d<6$. In Fig. 1 we sketch the entire
available range $2 < d < 6$, pointing out the various perturbative descriptions of the CFT's where $\epsilon$ expansions have been developed.

\begin{figure}[t]
\setcounter{figure}{0}
\includegraphics[width=16cm]{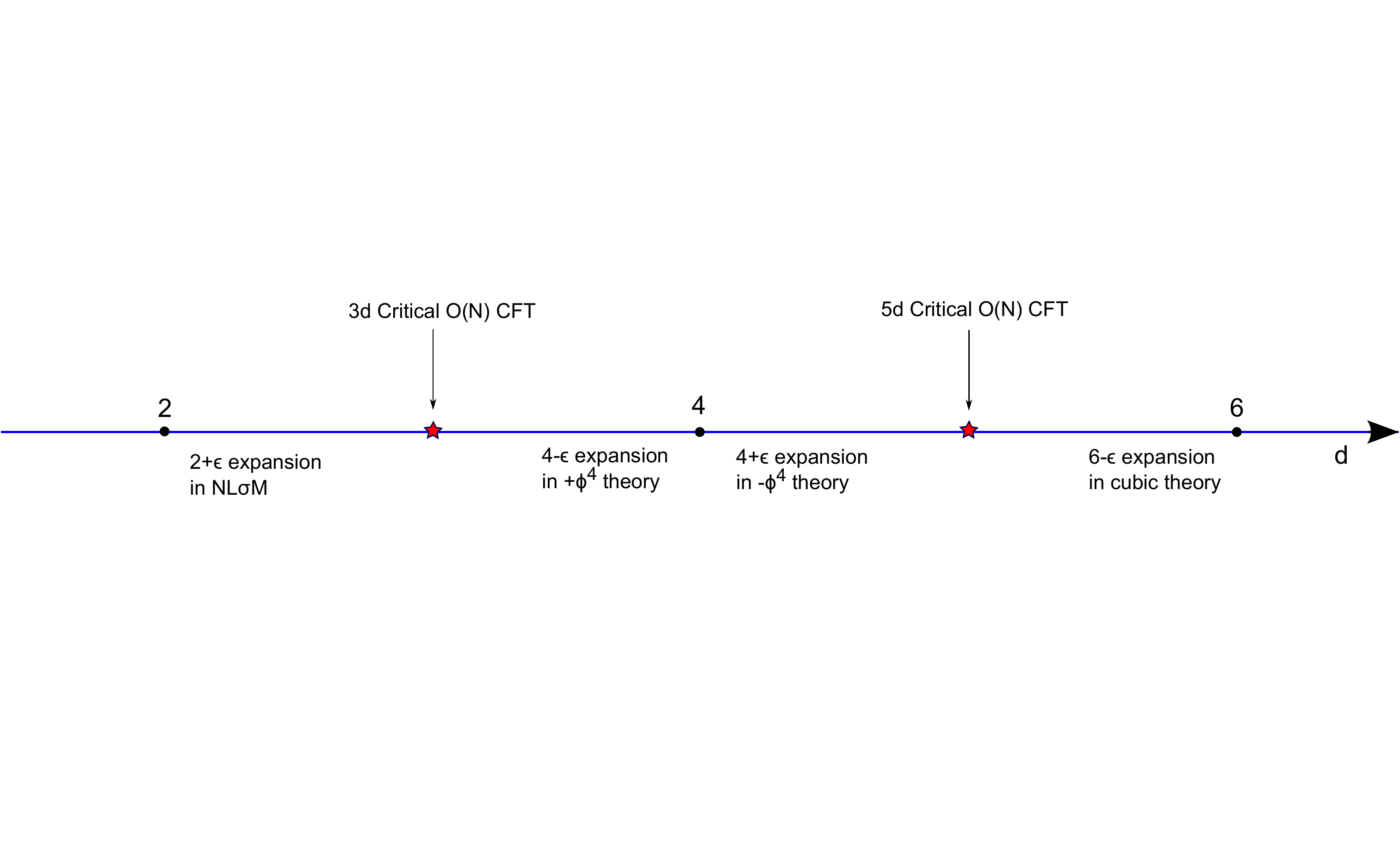}
\vskip -4cm
\caption{Interacting unitary $O(N)$ symmetric scalar CFT's exist for dimensions $2<d<6$, with $d=4$ excluded. In $6-\epsilon$ and $4-\epsilon$ dimensions they
may be described as weakly coupled IR fixed points of the cubic and quartic scalar theories, respectively. In $4+\epsilon$ and $2+\epsilon$ dimensions they are
weakly coupled UV fixed points of the quartic theory and of the $O(N)$ Non-linear $\sigma$ Model, respectively.}
\label{fig1}
\end{figure}

In Section \ref{cttheorem} we discuss the large $N$ results for $C_T$, the coefficient of the two-point function of the stress-energy tensor \cite{Petkou:1995vu}. We note that, as
$d$ approaches 6, $C_T$ approaches that of the free theory of $N+1$ scalar fields. This gives further evidence for our proposal that the IR fixed point of
(\ref{cubic6d}) describes the $O(N)$ symmetric CFT. We show that the RG flow from this interacting CFT to the free theory of $N$ scalars provides a counter example
to the conjectured $C_T$ theorem. On the other hand, the five-dimensional version of the $F$-theorem \cite{Klebanov:2011gs} holds for this RG flow.

Our discussion of the $(\phi^i \phi^i)^2$ scalar theory in the range $4<d<6$ is analogous to the much earlier results \cite{Hasenfratz:1991it,ZinnJustin:1991yn,Moshe:2003xn} about the Gross-Neveu model \cite{Gross:1974jv} in the range
$2<d<4$. The latter is a $U(\tilde N)$ invariant theory of $\tilde N$ Dirac fermions with an irrelevant quartic interaction $(\bar \psi^i \psi^i)^2$. Using the Hubbard-Stratonovich transformation,
it is not hard to show that this model has a UV fixed point, at least for large $\tilde N$ \cite{Moshe:2003xn}. This CFT was conjectured \cite{Sezgin:2003pt,Leigh:2003gk} to be dual to the type B Vasiliev theory in AdS$_4$ with appropriate boundary conditions. An alternative, UV complete description of this CFT is via the Gross-Neveu-Yukawa (GNY) model \cite{Hasenfratz:1991it,ZinnJustin:1991yn,Moshe:2003xn}, which is a theory of $\tilde N$ Dirac fermions coupled to a scalar field $\sigma$ with $U(\tilde N)$ invariant interactions $g_1 \sigma  \bar \psi^i \psi^i + g_2 \sigma^4/24$. This description is weakly coupled in $d=4-\epsilon$, and it is not hard to show that the one-loop beta functions have an IR stable fixed point for any positive $\tilde N$. The operator dimensions
at this fixed point match the large $\tilde N$ treatment of the Gross-Neveu model \cite{Derkachov:1993uw}.
These results will be reviewed in Section \ref{GNYukawa}, where
we also discuss tests of the 3-d F-theorem \cite{Jafferis:2011zi,Klebanov:2011gs} provided by the GNY model.

\section{Review of large $N$ results for the critical $O(N)$ CFT}
\label{largeNreview}

Let us consider the Euclidean field theory of $N$ real massless scalar fields with an $O(N)$ invariant quartic interaction
\begin{equation}
S = \int d^dx\left(\frac{1}{2}(\partial \phi^i )^2 +\frac{\lambda}{4}(\phi^i\phi^i)^2\right)\,.
\label{quartic}
\end{equation}
As follows from dimensional analysis, the interaction term is relevant for $d<4$ and irrelevant for $d>4$. Hence, for $2<d<4$, it is expected that the quartic interaction
generates a flow from the free UV fixed point to an interacting IR fixed point. In $d=4-\epsilon$ this fixed point can be studied perturbatively in the framework of the Wilson-Fisher $\epsilon$-expansion \cite{Wilson:1971dc,Wilson:1973jj}. Indeed, the one-loop beta function for the theory in $d=4-\epsilon$ reads
\begin{equation}
\beta_{\lambda} = -\epsilon \lambda + (N+8)\frac{\lambda^2}{8\pi^2}
\ ,\end{equation}
and there is a weakly coupled IR fixed point at
\begin{equation}
\lambda_* = \frac{8\pi^2}{N+8}\epsilon\,.
\end{equation}
Higher order corrections in $\epsilon$ will change the value of the critical coupling, but not its existence, at least in perturbation theory.
The anomalous dimensions of the fundamental field $\phi^i$ and the composite $\phi^i\phi^i$ at the fixed point can be computed to be, to leading order in $\epsilon$
\begin{equation}
\gamma_{\phi} = \frac{N+2}{4(N+8)^2}\epsilon^2+{\cal O}(\epsilon^3)\,,\qquad \gamma_{\phi^2} = \frac{N+2}{N+8}\epsilon +{\cal O}(\epsilon^2)
\ ,\label{anom-epsilon}
\end{equation}
corresponding to the scaling dimensions
\begin{eqnarray}
\label{delta-phi}
\Delta_{\phi} &=& \frac{d}{2}-1+\gamma_{\phi} = 1-\frac{\epsilon}{2} +\frac{N+2}{4(N+8)^2}\epsilon^2+{\cal O}(\epsilon^3)\ ,\\
\Delta_{\phi^2} &=& d-2+\gamma_{\phi^2}=2-\frac{6}{N+8}\epsilon+{\cal O}(\epsilon^2)\,.
\label{delta-sigma}
\end{eqnarray}
Note that the dimension of the $\phi^i\phi^i$ operator is $2+{\cal O}(1/N)$ to leading order at large $N$, a result that follows from the large $N$ analysis reviewed below. Higher order corrections in $\epsilon$ for general $N$ may be derived by higher loop calculations in the theory (\ref{quartic}), and they are known up to order $\epsilon^5$ \cite{Kleinert:1991rg, Guida:1998bx}.

For $d>4$, the interaction is irrelevant and so the IR fixed point is the free theory; however, one may ask about the existence of interacting UV fixed points. Working in $d=4+\epsilon$ for small $\epsilon$, one indeed finds a perturbative UV fixed point at (see, for example \cite{Giombi:2014iua})
\begin{equation}
\lambda_* = -\frac{8\pi^2}{N+8}\epsilon\,.
\end{equation}
The anomalous dimensions at this critical point are given by the same expressions (\ref{anom-epsilon}) with $\epsilon \rightarrow -\epsilon$. Note that, because $\gamma_{\phi}$ starts at order $\epsilon^2$, the dimension of $\phi$ stays above the unitarity bound for all $N$, at least for sufficiently small $\epsilon$. However, since the fixed point requires a negative coupling, one may worry about its stability, and it is important to study this critical point by alternative methods.

A complementary approach to the $\epsilon$ expansion that can be developed at arbitrary dimension $d$ is the large $N$ expansion. The standard technique to study the theory (\ref{quartic}) at large $N$ is based on introducing a Hubbard-Stratonovich auxiliary field $\sigma$ as
\begin{equation}
S = \int d^d x \left(\frac{1}{2}(\partial \phi^i)^2+\frac{1}{2}\sigma \phi^i\phi^i-\frac{\sigma^2}{4\lambda}\right)\,.
\label{hubbard}
\end{equation}
Integrating out $\sigma$ via its equation of motion $\sigma = \lambda \phi^i\phi^i$, one gets back to the original lagrangian. The quartic interaction in (\ref{quartic}) may in fact be viewed as a particular example of the double trace deformations studied in \cite{Gubser:2002vv}. One can then show that at large $N$ the dimension of $\phi^i\phi^i$ goes from $\Delta=d-2$ at the free fixed point to $d-\Delta=2$ at the interacting fixed point. At the conformal point, the last term in (\ref{hubbard}) can be dropped\footnote{This applies formally to both IR and UV fixed points.}, and the field $\sigma$ plays the role of the composite operator $\phi^i\phi^i$. One may then study the critical theory using the action
\begin{equation}
S_{\rm crit} = \int d^dx \left(\frac{1}{2}(\partial \phi^i)^2+\frac{1}{2\sqrt{N}}\sigma \phi^i\phi^i\right)
\label{S-crit}
\end{equation}
where we have rescaled $\sigma$ by a factor of $\sqrt{N}$ for reasons that will become clear momentarily. The $1/N$ perturbation theory can be developed by integrating out the fundamental fields $\phi^i$. This generates an effective non-local kinetic term for $\sigma$
\begin{equation}
\begin{aligned}
Z &= \int D\phi D\sigma\, e^{-\int d^d x\left(\frac{1}{2}(\partial \phi^i)^2+\frac{1}{2\sqrt{N}}\sigma \phi^i\phi^i\right)}\\
&=\int D\sigma\, e^{\frac{1}{8N}\int d^d x d^d y\, \sigma(x)\sigma(y)\, \langle \phi^i\phi^i(x)\phi^j\phi^j(y)\rangle_0+{\cal O}(\sigma^3)}
\end{aligned}
\label{F-sigma}
\end{equation}
where we have assumed large $N$ and the subscript `$0$' denotes expectation values in the free theory. We have
\begin{equation}
\langle \phi^i\phi^i(x)\phi^j\phi^j(y)\rangle_0 = 2N [G(x-y)]^2\,,\qquad G(x-y)= \int \frac{d^dp}{(2\pi)^d} \frac{e^{ip(x-y)}}{p^2}\,.
\end{equation}
In momentum space, the square of the $\phi$ propagator reads
\begin{eqnarray}
&&[G(x-y)]^2 = \int \frac{d^d p}{(2\pi)^d} e^{ip(x-y)}\tilde G(p)\cr
&&\tilde G(p)=\int \frac{d^dq}{(2\pi)^d} \frac{1}{q^2(p-q)^2} = -\frac{(p^2)^{d/2-2}}{2^d (4\pi)^{\frac{d-3}{2}}\Gamma(\frac{d-1}{2})\sin(\frac{\pi d}{2})}
\end{eqnarray}
and so from (\ref{F-sigma}) one finds the two-point function of $\sigma$ in momentum space
\begin{equation}
\langle \sigma(p)\sigma(-p)\rangle = 2^{d+1} (4\pi)^{\frac{d-3}{2}}\Gamma\left(\frac{d-1}{2}\right)\sin(\frac{\pi d}{2})\, (p^2)^{2-\frac{d}{2}} \equiv \tilde C_{\sigma}  (p^2)^{2-\frac{d}{2}}\,.
\label{sigma-prop}
\end{equation}
The corresponding two-point function in coordinate space can be obtained by Fourier transform and reads
\begin{equation}
\langle \sigma(x)\sigma(y) \rangle = \frac{2^{d+2}\Gamma\left(\frac{d-1}{2}\right)\sin(\frac{\pi d}{2})}{\pi^{\frac{3}{2}}\Gamma\left(\frac{d}{2}-2\right)}\,\frac{1}{|x-y|^4}
\equiv \frac{C_{\sigma}}{|x-y|^4}\,.
\label{sigma-prop-x}
\end{equation}
Indeed, this is the two-point function of a conformal scalar operator of dimension $\Delta=2$. Note also that the coefficient $C_{\sigma}$ is positive in the range $2<d<6$.

The large $N$ perturbation theory can then be developed using the propagator (\ref{sigma-prop})-(\ref{sigma-prop-x}) for $\sigma$, the canonical propagator for $\phi$ and the interaction term $\sigma\phi^i\phi^i$ in (\ref{S-crit}). For instance, the $1/N$ term in the anomalous dimension of $\phi^i$ can be computed from the one-loop correction to the $\phi$-propagator
\begin{eqnarray}
\frac{1}{N} \int \frac{d^dq}{(2\pi)^d} \frac{1}{(p-q)^2} \frac{\tilde C_{\sigma}}{(q^2)^{\frac{d}{2}-2+\delta}}\,,
\end{eqnarray}
where we have introduced a small correction $\delta$ to the power of the $\sigma$ propagator as a regulator\footnote{One may perform the calculation using a momentum cutoff, yielding the same final result.}. Doing the momentum integral by using (\ref{Iab}), one obtains the result
\begin{equation}
\frac{\tilde C_{\sigma}}{N} \frac{(d-4)}{(4\pi)^{\frac{d}{2}}d \Gamma(\frac{d}{2})} (p^2)^{1-\delta}\Gamma(\delta-1)
\end{equation}
where we have set $\delta=0$ in the irrelevant factors. The $1/\delta$ pole corresponds to a logarithmic divergence and is cancelled as usual by the wave function renormalization of $\phi$. Defining the dimension of $\phi$ as
\begin{equation}
\Delta_{\phi} = \frac{d}{2}-1+\frac{1}{N}\eta_1+\frac{1}{N^2}\eta_2+\ldots
\label{delphi-conv}
\end{equation}
the one-loop calculation above yields the result
\begin{equation}
\eta_1  = \frac{\tilde C_{\sigma}(d-4)}{(4\pi)^{\frac{d}{2}} d \Gamma(\frac{d}{2})} = \frac{2^{d-3}(d-4)\Gamma\left(\frac{d-1}{2}\right)\sin\left(\frac{\pi d}{2}\right)}{\pi^{\frac{3}{2}} \Gamma\left(\frac{d}{2}+1\right)}\,.
\label{eta1-phi}
\end{equation}
Setting $d=4-\epsilon$ and expanding for small $\epsilon$, it is straightforward to check that this agrees with the $1/N$ term of (\ref{delta-phi}). The leading anomalous dimension of $\sigma$ also takes a simple form \cite{Vasiliev:1981yc,Petkou:1994ad,Petkou:1995vu}
\begin{equation}
\Delta_{\sigma}=2+\frac{1}{N}\frac{4(d-1)(d-2)}{d-4}\eta_1+{\cal O}(\frac{1}{N^2})\,.
\end{equation}
and can be seen to precisely agree with (\ref{delta-sigma}) in $d=4-\epsilon$.

\begin{figure}
\centering
\parbox{5cm}{
\includegraphics[width=6cm]{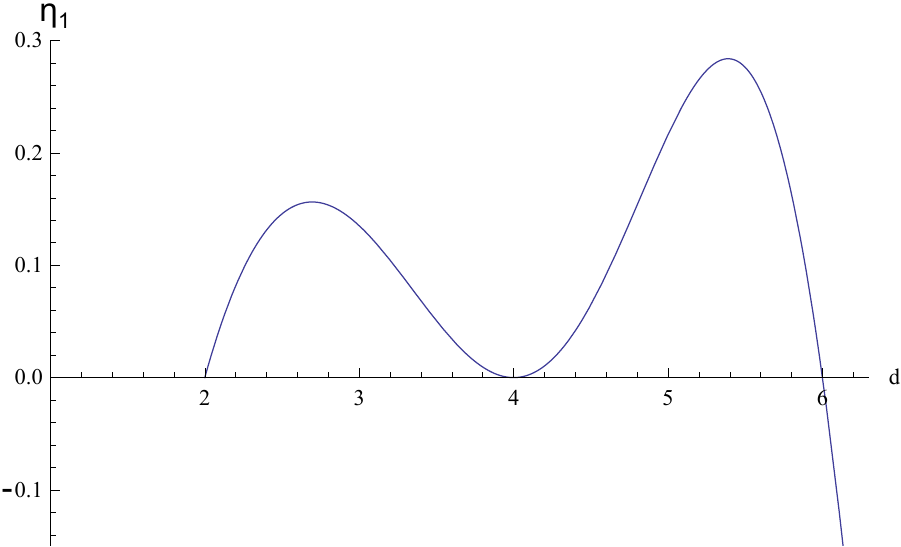}}
\qquad \qquad \qquad
\begin{minipage}{5cm}
\includegraphics[width=6cm]{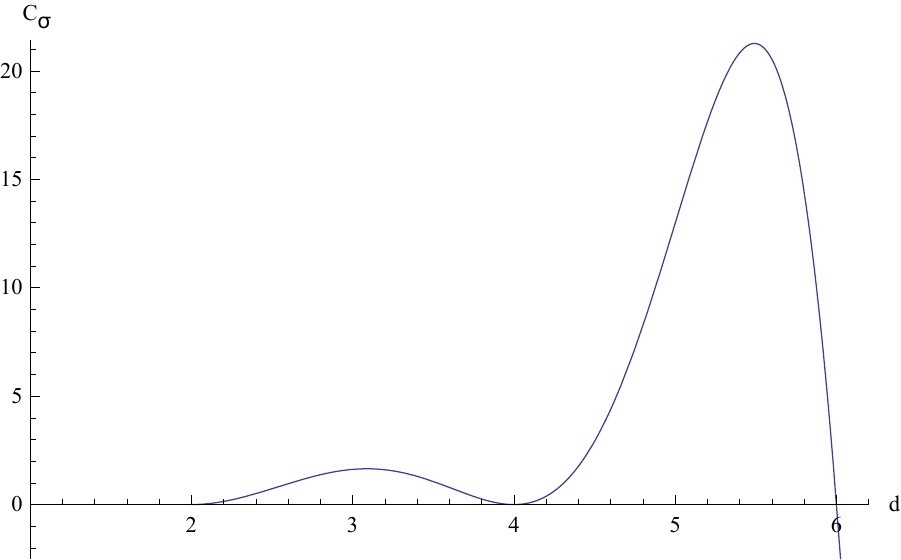}
\end{minipage}
\caption{The $1/N$ anomalous dimension of $\phi^i$ and the coefficient of the two-point function of $\sigma$ in the large $N$ critical $O(N)$ theory for $2<d<6$.}
\label{eta1}
\end{figure}

Note that the anomalous dimension of $\phi^i$ (\ref{eta1-phi}) is positive for all $2<d<6$. Thus, while the focus of most of the exisiting literature is on the range $2<d<4$, we see no obvious problems with unitarity in continuing the large $N$ critical $O(N)$ theory to the range $4<d<6$.\footnote{Recall that the unitarity bound for a scalar operator is $\Delta \ge d/2-1$. One can also check that the order $1/N$ term in the anomalous dimension of the $O(N)$ invariant higher spin currents, given in \cite{Lang:1992zw}, is positive in the range $2<d<6$, consistently with the unitarity bound $\Delta_s \ge s+d-2$ for spin $s$ operators ($s>1/2$).} A plot of $\eta_1$ and of the $\sigma$ two-point function coefficient $C_{\sigma}$ is given in Figure \ref{eta1}, showing that they are both positive for $2<d<6$.

At higher orders in the $1/N$ expansion, a straightforward diagrammatic approach becomes rather cumbersome. However, the conformal bootstrap method developed in
papers by A.~N.~Vasiliev and collaborators \cite{Vasiliev:1981yc,Vasiliev:1981dg,Vasiliev:1982dc} has allowed to compute the anomalous dimension of $\phi^i$ to order $1/N^3$ and that of $\sigma$ to order $1/N^2$. These results have been successfully matched to all available orders in the $d=2+\epsilon$ and $d=4-\epsilon$ expansions, providing a strong test of their correctness. The explicit form of $\eta_2$ in general dimensions, defined as in (\ref{delphi-conv}), reads \cite{Vasiliev:1981yc,Vasiliev:1981dg}:\footnote{Ref. \cite{Vasiliev:1981dg} contains an apparent typo: in the definition of the function $v'(\mu)$ given in Eq. (21), $\alpha=\mu-1$ should be replaced by $\alpha=\mu-2$. The correct formula for
$\eta_2$ may be found in the earlier paper \cite{Vasiliev:1981yc}.}
\begin{eqnarray}
\!\!\!\!\!\!\!\!\!\!&&\eta_2 = 2\eta_1^2\left(f_1+f_2+f_3\right)\ ;\nonumber \\
\!\!\!\!\!\!\!\!\!\!&&f_1 = v'(\mu)+\frac{\mu^2+\mu-1}{2\mu(\mu-1)}\,,\quad f_2 =\frac{\mu}{2-\mu}v'(\mu) +\frac{\mu(3-\mu)}{2(2-\mu)^2}\,, \quad
f_3 = \frac{\mu(2\mu-3)}{2-\mu}v'(\mu)+\frac{2\mu(\mu-1)}{2-\mu}\ ;\cr
\!\!\!\!\!\!\!\!\!\!&&v'(\mu) = \psi(2-\mu)+\psi(2\mu-2)-\psi(\mu-2)-\psi(2)\,,\qquad \psi(x)=\frac{\Gamma'(x)}{\Gamma(x)}\,,\qquad \mu=\frac{d}{2}\,.
\end{eqnarray}
The expressions for the dimension of $\sigma$ at order $1/N^2$ \cite{Vasiliev:1981dg} and for the coefficient $\eta_3$ \cite{Vasiliev:1982dc} in arbitrary dimensions are lengthy and we do not report them explicitly here.\footnote{Note that \cite{Vasiliev:1981dg} derives a result for the critical exponent $\nu$, which is related to the dimension of $\sigma$ by $\Delta_{\sigma} = d-\frac{1}{\nu}$. We also note that a misprint in eq. (22) of \cite{Vasiliev:1982dc} has been later corrected in eq. (11) of \cite{Derkachov:1993uw}.} However, since they are useful to test our general picture,
we write below the explicit $\epsilon$-expansions of $\Delta_{\phi}$ and $\Delta_{\sigma}$ in $d=6-\epsilon$, including all known terms in the $1/N$ expansion:
\begin{eqnarray}
\!\!\!\!\!\Delta_{\phi} = 2-\frac{\epsilon}{2}+\left(\frac{1}{N}+\frac{44}{N^2}+\frac{1936}{N^3}+\ldots\right)\epsilon
-\left(\frac{11}{12N}+\frac{835}{6N^2}+\frac{16352}{N^3}+\ldots \right)\epsilon^2+{\cal O}(\epsilon^3)
\label{delphi-6mep}
\end{eqnarray}
and
\begin{eqnarray}
\Delta_{\sigma}= 2+\left(\frac{40}{N}+\frac{6800}{N^2}+\ldots\right)\epsilon
-\left(\frac{104}{3N}+\frac{34190}{3N^2}+\ldots \right)\epsilon^2+{\cal O}(\epsilon^3)\ .
\label{delsigma-6mep}
\end{eqnarray}
In the next section, we will show that the order $\epsilon$ terms precisely match the one-loop anomalous dimensions at the IR fixed point of the cubic theory
(\ref{cubic6d}) in $d=6-\epsilon$. Higher orders in $\epsilon$ should be compared to higher loop contributions in the $d=6-\epsilon$ cubic theory, and it would be interesting to match them as well.

Using the results in \cite{Vasiliev:1981yc,Vasiliev:1981dg,Vasiliev:1982dc, Derkachov:1993uw} we may also compute the dimension of $\phi^i$ and $\sigma$ directly in the physical dimension $d=5$. The term of order $1/N^3$ depends on a non-trivial self-energy integral that was not evaluated for general dimension in \cite{Vasiliev:1982dc}. An explicit derivation of this integral in general $d$ was later obtained in \cite{Broadhurst:1996yc}. Using that result, we find in $d=5$
\begin{eqnarray}
\Delta_{\phi} &=& \frac{3}{2}+\frac{32}{15\pi^2N}-\frac{1427456}{3375 \pi^4N^2} \cr
&+&\left(\frac{275255197696}{759375 \pi^6}-\frac{89735168}{2025 \pi^4}+\frac{32768 \ln 4}{9 \pi^4}-\frac{229376 \zeta(3)}{3\pi^6}\right)\frac{1}{N^3}+\ldots \cr
&=& \frac{3}{2}+\frac{0.216152}{N}-\frac{4.342}{N^2}-\frac{121.673}{N^3}+\ldots
\label{delphi-5d}
\end{eqnarray}
and
\begin{eqnarray}
\Delta_{\sigma} =2+ \frac{512}{5 \pi^2 N}+\frac{2048 \left(12625\pi^2-113552\right)}{1125 \pi^4 N^2}+\ldots
=2+\frac{10.3753}{N}+\frac{206.542}{N^2}+\ldots
\end{eqnarray}

We note that the coefficients of the $1/N$ expansion are considerably larger than in the $d=3$ case.\footnote{In $d=3$, one finds \cite{Vasiliev:1981dg, Vasiliev:1981yc, Vasiliev:1982dc} (see also \cite{Kos:2013tga}) $\Delta_{\phi} = \frac{1}{2}+\frac{0.135095}{N}-\frac{0.0973367}{N^2}-\frac{0.940617}{N^3}+{\cal O}(1/N^4)$ and $\Delta_{\sigma} = 2-\frac{1.08076}{N}-\frac{3.0476}{N^2}+{\cal O}(1/N^3)$.} Assuming the result (\ref{delphi-5d}) to order $1/N^3$, one finds that the dimension of $\phi^i$ goes below unitarity at $N_{\rm crit} = 35$. This is much lower than the value $N_{\rm crit}=1038$ that we will find in $d=6-\epsilon$. This is just a rough estimate, since the $1/N$ expansion is only asymptotic and should be analyzed with care. Note for instance that in the $d=3$ case, a similar estimate to order $1/N^3$ would suggest a critical value $N_{\rm crit} = 3 $, while in fact there is no lower bound: at $N=1$ we have the 3d Ising model, where it is known \cite{2010PhRvB..82q4433H,ElShowk:2012ht,Kos:2013tga,El-Showk:2014dwa} that $\Delta_{\phi}
\approx 0.518 > 1/2$. Nevertheless, the reduction from a very large value $N_{\rm crit}=1038$ in $d=6-\epsilon$ to a much smaller critical value in $d=5$ is not unexpected.
For instance, an analogous phenomenon is known to occur in the Abelian Higgs model containing $N_f$ complex scalars. A fixed point in $d=4-\epsilon$ exists only for
$N_f \geq 183$ \cite{Halperin:1973jh,Moshe:2003xn}, while non-perturbative studies directly in $d=3$ suggest a much lower critical value of $N_f$. \footnote{We thank D.~T.~Son for bringing this to our attention.} Some evidence for the reduction in the critical value of $N_f$ as $d$ is decreased comes from calculations
of higher order corrections in $\epsilon$
\cite{herbut1997herbut}. It would be nice to study such corrections for the $O(N)$ theory in $6-\epsilon$ dimensions.
It would also be very interesting to explore the 5d fixed point at finite $N$ by numerical bootstrap methods similar to what has been done in $d=3$ in \cite{Kos:2013tga}.
For the non-unitary theory with $N=0$ such bootstrap studies were carried out very recently \cite{Gliozzi:2014jsa}.\footnote{
After the original version of this paper appeared, a bootstrap study of the $O(N)$ model in $d=5$ was carried out in \cite{Nakayama:2014yia}
with very encouraging results.}

The large $N$ critical $O(N)$ theory in general $d$ was further studied in a series of works by Lang and Ruhl \cite{Lang:1990ni, Lang:1991kp, Lang:1992pp, Lang:1992zw} and Petkou \cite{Petkou:1994ad,Petkou:1995vu}. Using conformal symmetry and self-consistency of the OPE expansion, various results about the operator spectrum of the critical theory were derived. As an example of interest to us, \cite{Lang:1992zw} derived an explicit formula for the anomalous dimension of the operator $\sigma^k$ (the $k$-th power of the auxiliary field), which reads
\begin{equation}
\Delta(\sigma^k) = 2k -\frac{2k (d-1)  \left((k-1)d^2 +d+4-3kd\right)}{d-4}\frac{\eta_1}{N}+{\cal O}(\frac{1}{N^2})
\end{equation}
where $\eta_1$ is the $1/N$ anomalous dimension of $\phi^i$ given in (\ref{eta1-phi}). In $d=6-\epsilon$, this gives
\begin{equation}
\label{primaries}
\Delta(\sigma^k) = 2k +(130k-90k^2)\frac{\epsilon}{N}+{\cal O}(\epsilon^2)\,.
\end{equation}
For $k=2,3$, we will be able to match this result with the one-loop operator mixing calculations in the cubic theory (\ref{cubic6d}).

In \cite{Petkou:1994ad,Petkou:1995vu}, explicit results for the 3-point function coefficients $g_{\phi\phi \sigma}$ and $g_{\sigma^3}$ were also derived. These are defined by the correlation functions
\begin{equation}
\begin{aligned}
&\langle \phi^i(x_1)\phi^j(x_2)\sigma(x_3)\rangle = \frac{g_{\phi\phi\sigma}}{|x_{12}|^{2\Delta_{\phi}-\Delta_{\sigma}}|x_{23}|^{\Delta_{\sigma}}|x_{13}|^{\Delta_{\sigma}}}\delta^{ij}\,.\\
&\langle \sigma(x_1)\sigma(x_2)\sigma(x_3)\rangle = \frac{g_{\sigma^3}}{\left(|x_{12}||x_{23}||x_{13}|\right)^{\Delta_{\sigma}}}
\end{aligned}
\end{equation}
The coefficient $g_{\phi\phi \sigma}$ was given in \cite{Petkou:1994ad,Petkou:1995vu} to order $1/N^2$ and arbitrary $d$. Expanding that result to leading order in $\epsilon = 6-d$, we find
\begin{equation}
g_{\phi\phi \sigma}^2 = \frac{6\epsilon}{N}\left(1+\frac{44}{N}+{\cal O}(\frac{1}{N^2})\right)\,.\\
\label{gphiphisigma}
\end{equation}
This result indeed matches the value of the coupling $g_1^2$ in (\ref{cubic6d}) at the IR fixed point, as will be shown in the next section. To leading order in $1/N$, the 3-point function coefficient $g_{\sigma^3}$ is related to $g_{\phi\phi\sigma}$ by \cite{Petkou:1994ad}
\begin{equation}
g_{\sigma^3} = 2(d-3)g_{\phi\phi \sigma}\,,
\label{g2}
\end{equation}
which, as we will see, is precisely consistent with the ratio $\frac{g_2^*}{g_1^*} = 6+{\cal O}(1/N)$ of the coupling constants at the $d=6-\epsilon$ IR fixed point.

\section{The IR fixed point of the cubic theory in $d=6-\epsilon$}
\label{IRfixed}

\begin{figure}[h]
\includegraphics[width=16cm]{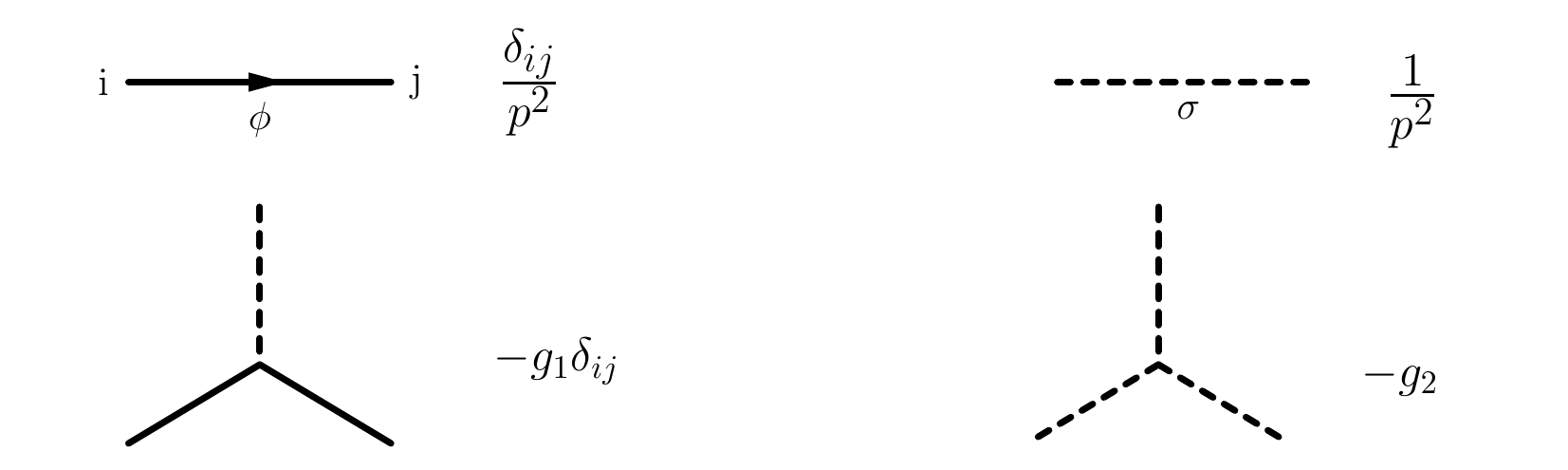}
\caption{Feynman rules of the theory in Euclidean space.}
\label{fig3}
\end{figure}

In this section, we show that the interacting scalar theory  with Lagrangian (\ref{cubic6d}) has a perturbative large $N$ IR fixed point in $d=6-\epsilon$, and compute the anomalous
dimensions of the fundamental fields $\phi^i$ and $\sigma$ at the fixed point. The theory (\ref{cubic6d}) has an obvious $O(N)$ symmetry, with the $N$-component vector $\phi^i, i=1,\ldots,N$ transforming in the fundamental representation. Note that $g_1$ and $g_2$ are classically marginal in $d=6$, and have dimension $\frac{\epsilon}{2}$ in $d=6-\epsilon$. The Feynman rules of the theory are shown in Figure \ref{fig3}.

\begin{figure}[h]
\centering
\includegraphics{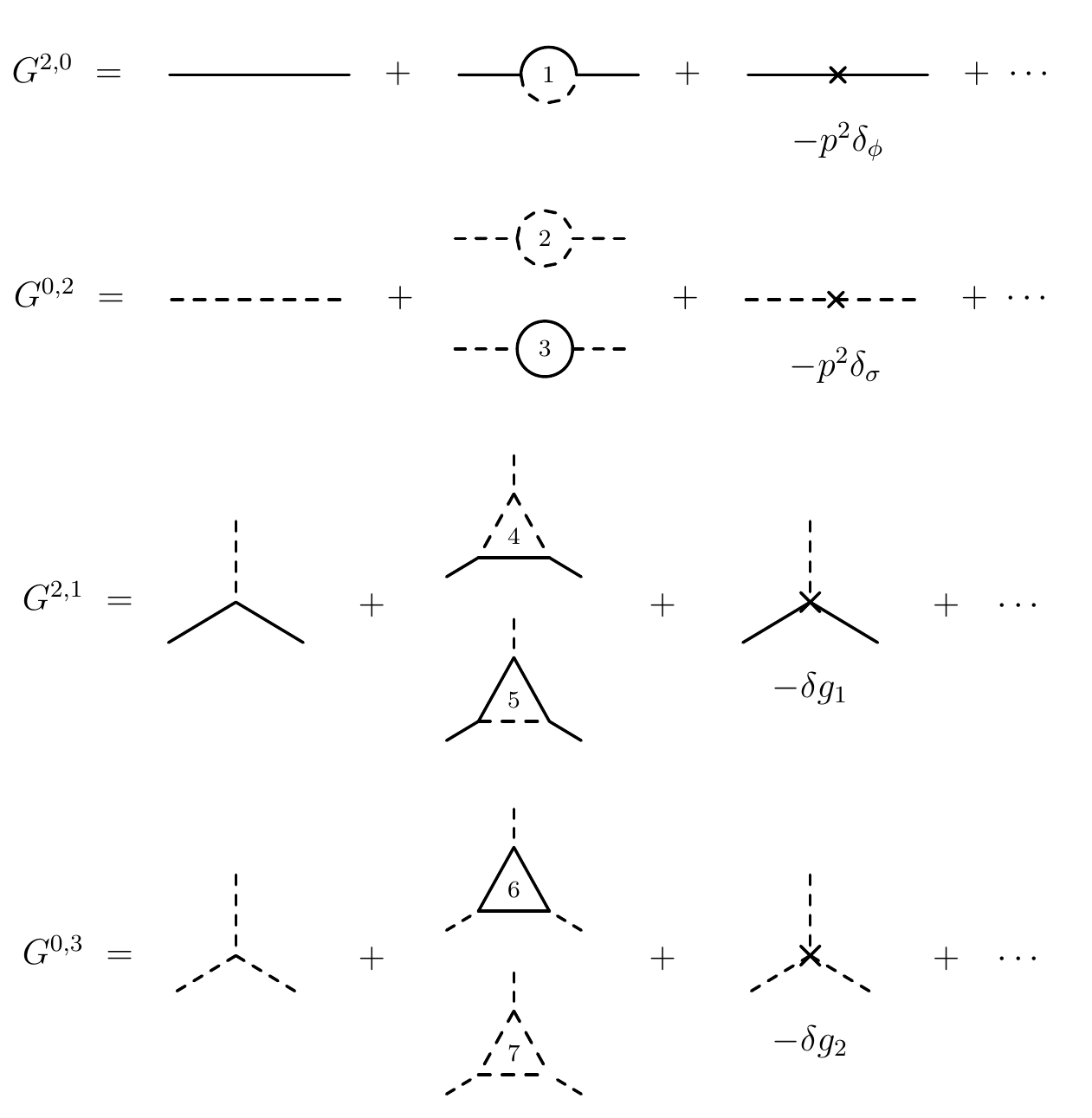}
\caption{Diagrams contributing to the 1-loop $\beta$-functions.}
\label{fig4}
\end{figure}

It is not hard to compute the one-loop beta functions $\beta_1$, $\beta_2$ for the couplings $g_1,g_2$. The relevant one-loop diagrams needed to compute the counter terms $\delta_{\phi}$, $\delta_{\sigma}$, $\delta g_1$ for the $\sigma \phi \phi$ coupling, and $\delta g_2$ for the $\sigma^3$ coupling are given in Figure \ref{fig4}. Note that $G^{m,n}$ denote the Green's function with $m$ $\phi$ fields and $n$ $\sigma$ fields. The one loop diagrams are labeled 1 through 7 for convenience.

Let us begin with the computation of diagram 1 in Figure \ref{fig4} \footnote{We will state approximate expressions for the one-loop integrals $I_1$ and $I_2$
that are sufficient for
extracting the $\log M^2$ terms in $d=6-\epsilon$. The more precise expressions are given in Appendix A.}
\begin{align}
D_1 = (-g_1)^2 \int \frac{d^dk}{(2\pi)^d}\frac{1}{(p+k)^2}\frac{1}{k^2} = -g_1^2 I_1 = -\frac{p^2}{(4\pi)^{3}} \frac{g_1^2}{6} \frac{\Gamma(3-d/2)}{(M^2)^{3-d/2}}\ .
\end{align}
Here we have used the renormalization condition $p^2=M^2$.
This has a $1/\epsilon$ pole in $d=6-\epsilon$ which must be canceled by the counter term $-p^2 \delta_{\phi}$. So we get:
\begin{equation}
\delta_{\phi} = -\frac{1}{(4\pi)^{3}} \frac{g_1^2}{6} \frac{\Gamma(3-d/2)}{(M^2)^{3-d/2}} \,.
\label{deltaz1}
\end{equation}

\vskip 1cm
For $\delta_{\sigma}$, we have two one-loop diagrams (2 and 3). However, other than having different coupling constant factors, the integrals are identical to diagram 1. Note that these two diagrams have a symmetry factor of 2, and there is a factor of $N$ associated with the $\phi$ loop in diagram 3. So, we have:
\begin{align}
D_2 + D_3 &= \frac{1}{2}(-g_2)^2 I_1 + \frac{N}{2}(-g_1)^2 I_1 = -\frac{Ng_1^2+g_2^2}{2}I_1\,.
\end{align}
We arrive at the following expression for $\delta_{\sigma}$:
\begin{equation}
\delta_{\sigma} = -\frac{1}{(4\pi)^{3}} \frac{Ng_1^2+g_2^2}{12} \frac{\Gamma(3-d/2)}{(M^2)^{3-d/2}}\,.
\label{deltaz2}
\end{equation}

Now let us compute corrections to the 3-point functions. Diagram 4 gives:
\begin{align}
D_4 &= (-g_1)^2 (-g_2) \int \frac{d^dk}{(2\pi)^d}\frac{1}{(k-p)^2}\frac{1}{(k+q)^2}\frac{1}{k^2} \nonumber\\
&= -g_1^2 g_2 I_2 = \frac{- g_1^2 g_2}{2(4\pi)^{3}}\frac{\Gamma(3-d/2)}{(M^2)^{3-d/2}}\,,
\end{align}
where $I_2$ is computed in Appendix A.
Diagram 5 is again exactly the same as diagram 4, except for the coupling factors:
\begin{align}
D_5 &= (-g_1)^3 \int \frac{d^dk}{(2\pi)^d}\frac{1}{(k-p)^2}\frac{1}{(k+q)^2}\frac{1}{k^2} \nonumber \\
&= - g_1^3 I_2 = \frac{- g_1^3}{2(4\pi)^{3}}\frac{\Gamma(3-d/2)}{(M^2)^{3-d/2}}\,.
\end{align}
The divergences in $D_4$ and $D_5$ must be canceled by the $-\delta g_1$ counterterm. So we get
\begin{equation}
\delta g_1 = -\frac{g_1^3+g_1^2 g_2}{2(4\pi)^{3}}\frac{\Gamma(3-d/2)}{(M^2)^{3-d/2}}\,.
\end{equation}

Finally, we calculate $\delta g_2$. The diagrams have the same topology, and just differ in the coupling factors. Also, in diagram 6, we have a factor of $N$ from the $\phi$ loop. So we find
\begin{align}
D_6 + D_7 &= N(-g_1)^3 I_2 + (-g_2)^3 I_2 \nonumber \\
&= - (Ng_1^3 + g_2^3) I_2 = \frac{- (Ng_1^3 +g_2^3)}{2(4\pi)^{3}}\frac{\Gamma(3-d/2)}{(M^2)^{3-d/2}}\,.
\end{align}
This term is canceled by $-\delta g_2$, so we have
\begin{equation}
\delta g_2 = -\frac{Ng_1^3+g_2^3}{2(4\pi)^{3}}\frac{\Gamma(3-d/2)}{(M^2)^{3-d/2}}\,.
\end{equation}

The Callan-Symanzik equation for the Green's function $G^{m,n}$ is:
\begin{equation}
(M\frac{\partial}{\partial M} + \beta_1 \frac{\partial}{\partial g_1} + \beta_2 \frac{\partial}{\partial g_2} + m \gamma_{\phi} + n \gamma_{\sigma})G^{m,n} = 0
\end{equation}
Let's first apply it to $G^{2,0}$. Then, to leading order in perturbation theory, the Callan-Symanzik equation simplifies to
\begin{equation}
-\frac{1}{p^2}M\frac{\partial}{\partial M} \delta_{\phi} + 2 \gamma_{\phi} \frac{1}{p^2} = 0\,,
\end{equation}
which gives the anomalous dimension of $\phi^i$
\begin{align}
\gamma_{\phi} = \frac{1}{2}M\frac{\partial}{\partial M} \delta_{\phi}=\frac{1}{(4\pi)^{3}} \frac{g_1^2}{6}\,.
\label{g-phi}
\end{align}
Note that $\gamma_1 > 0$ as long as the coupling constant $g_1$ is real. Analogously, we obtain the $\sigma$ anomalous dimension
\begin{align}
\gamma_{\sigma} = \frac{1}{2}M\frac{\partial}{\partial M} \delta_{\sigma} =\frac{1}{(4\pi)^{3}} \frac{Ng_1^2+g_2^2}{12}\,.
\label{g-sigma}
\end{align}
Now we can compute the $\beta$ functions for $g_1$ and $g_2$.
We have
\begin{equation}
\beta_1 =-\frac{\epsilon}{2}g_1+ M\frac{\partial}{\partial M}(-\delta g_1 + \frac{1}{2} g_1 (2 \delta_{\phi}+ \delta_{\sigma}))\,,
\end{equation}
where the first term accounts for the bare dimension of $g_1$ in $d=6-\epsilon$. After some simple algebra, we obtain
\begin{equation}
\beta_1 = -\frac{\epsilon}{2}g_1+\frac{(N-8)g_1^3-12g_1^2 g_2 + g_1 g_2^2}{12(4\pi)^3}\,.
\label{beta1}
\end{equation}
Notice that when $N \gg 1$, $\beta_1$ is positive.

Finally, let us compute $\beta_2$. The Callan-Symanzik equation gives:
\begin{equation}
\beta_2 =-\frac{\epsilon}{2} g_2 +M\frac{\partial}{\partial M}(-\delta g_2 + \frac{1}{2} g_2 (3 \delta_{\sigma}))\ ,
\end{equation}
and so
\begin{equation}
\beta_2 =-\frac{\epsilon}{2}g_2+ \frac{-4Ng_1^3 +Ng_1^2 g_2 -3g_2^3}{4(4\pi)^3}\,.
\label{beta2}
\end{equation}
As a check, let us note that for $N=0$ the beta function for $g_2$ in $d=6$ reduces to
\begin{equation}
\beta_2 = -\frac{3g_2^3}{4(4\pi)^3}\,,
\label{sigma3beta}
\end{equation}
which is the correct result for the single scalar cubic field theory in $d=6$.

The single scalar cubic field theory in $d=6-\epsilon$ has no fixed points at real coupling, due to the negative sign of the beta function (\ref{sigma3beta}). It has a fixed point at imaginary coupling, which is conjectured to be related by dimensional continuation to the Yang-Lee edge singularity \cite{Fisher:1978pf}. However, as we now show, for sufficiently large $N$, our model has a stable interacting IR fixed point. Note that for large $N$ the beta functions simplify to
\begin{align}
\beta_1 &= -\frac{\epsilon}{2}g_1 + \frac{Ng_1^3}{12(4\pi)^3} \\
\beta_2 &= -\frac{\epsilon}{2}g_2 + \frac{-4Ng_1^3 +Ng_1^2 g_2}{4(4\pi)^3}\,.
\end{align}
This can be solved to get\footnote{There is also a physically equivalent solution with the opposite signs of $g_1, g_2$.}
\begin{align}
g_{1}^* &= \sqrt{\frac{6\epsilon (4\pi)^3}{N}}  \\
g_{2}^* &= 6 g_{1}^*\,.
\label{g26g1}
\end{align}
It is straightforward to compute the subleading corrections at large $N$ by solving the exact beta function equations (\ref{beta1}), (\ref{beta2}) in powers of $1/N$. This yields
\begin{align}
g_{1}^* &= \sqrt{\frac{6\epsilon (4\pi)^3}{N}}\left(1+\frac{22}{N}+\frac{726}{N^2}-\frac{326180}{N^3}+\ldots\right)\label{g1star}  \\
g_{2}^* &= 6 \sqrt{\frac{6\epsilon (4\pi)^3}{N}}\left(1+\frac{162}{N}+\frac{68766}{N^2}+\frac{41224420}{N^3}+\ldots\right)\label{g2star}\,.
\end{align}
Note that the coefficients in this expansion appear to increase quite rapidly. This suggests that the large $N$ expansion may break down at some finite $N$. Indeed, we will see in Section \ref{finiteN} that this large $N$ IR fixed point disappears at $N \leq 1038$ (the coupling constants go off to the complex plane). For all values of $N\geq 1039$, the fixed point has real couplings and is IR stable, see Section \ref{finiteN}. At large $N$, the IR stability of the fixed point can be seen from the fact that the matrix $\frac{\partial \beta_i}{\partial g_j}$ evaluated at the fixed point has two positive eigenvalues. These eigenvalues are in fact related to the dimensions of the two eigenstates coming from the operator mixing of $\sigma^3$ and $\sigma\phi^i\phi^i$ operators, as will be discussed in more detail in Section \ref{mixing6}.

We can now use the values of the fixed point couplings to compute the dimensions of the elementary fields $\phi^i$ and $\sigma$ in the IR. From (\ref{g-phi}) and (\ref{g-sigma}) we obtain
\begin{align}
\Delta_{\phi} &= \frac{d}{2}-1+\gamma_{\phi} = 2-\frac{\epsilon}{2}+\frac{1}{(4\pi)^{3}} \frac{(g_1^*)^2}{6}\nonumber \\
&=2-\frac{\epsilon}{2}+\frac{\epsilon}{N}+\frac{44\epsilon}{N^2}+\frac{1936\epsilon}{N^3}+\ldots
\label{delphi}
\end{align}
and
\begin{align}
\Delta_{\sigma} &= \frac{d}{2}-1+\gamma_{\sigma} = 2-\frac{\epsilon}{2}+\frac{1}{(4\pi)^{3}} \frac{N(g_1^*)^2+(g_2^*)^2}{12} \nonumber \\
&=2+\frac{40\epsilon}{N}+\frac{6800\epsilon}{N^2}+\ldots
\label{delsigma}
\end{align}
Note that both dimensions are above the unitarity bound in $d=6-\epsilon$, namely $\Delta > \frac{d}{2}-1$, since the anomalous dimensions are positive. Moreover, note that the order $\epsilon$ term in the dimension of $\sigma$ cancels out and we find $\Delta_{\sigma}=2+{\cal O}(1/N)$. This is in perfect agreement with the large $N$ description of the critical $O(N)$ CFT. In that approach, the field $\sigma$ corresponds to the composite operator $\phi^i\phi^i$, whose dimension goes from $\Delta=d-2$ at the free fixed point to $d-\Delta+{\cal O}(1/N)=2+{\cal O}(1/N)$ at the interacting fixed point. Furthermore, comparing (\ref{delphi}), (\ref{delsigma}) with (\ref{delphi-6mep}), (\ref{delsigma-6mep}), we see that the coefficients of the $1/N$ expansion precisely match the available results for the large $N$ critical $O(N)$ theory \cite{Vasiliev:1981dg,Vasiliev:1981yc,Vasiliev:1982dc} expanded at $d=6-\epsilon$. This is a strong test that the IR fixed point of the cubic theory in $d=6-\epsilon$ is identical at large $N$ to the dimensional continuation of the critical point of the $(\phi^i\phi^i)^2$ theory.

We may also match the values of the fixed point couplings (\ref{g1star}),(\ref{g2star}) with the large $N$ results (\ref{gphiphisigma}), (\ref{g2}) for the 3-point functions coefficients in the critical $O(N)$ theory. At leading order in $\epsilon$, the 3-point functions in our cubic model simply come from a tree level calculation, and it is straightforward to verify the agreement of (\ref{g1star}) with (\ref{gphiphisigma}),\footnote{An overall normalization factor comes from the normalization of the massless scalar propagators in $d=6$.} and that the ratio $g_1^*/g_2^* = 6$ at leading order at large $N$ agrees with (\ref{g2}).

\section{Analysis of fixed points at finite $N$}
\label{finiteN}

In this section we analyze the one-loop fixed points for general $N$. First, let us define:
\begin{align}
g_1 &\equiv \sqrt{\frac{6\epsilon (4\pi)^3}{N}} x\, ,\label{g1x}\\
g_2 &\equiv \sqrt{\frac{6\epsilon (4\pi)^3}{N}} y \,. \label{g2y}
\end{align}
After this, the vanishing of the $\beta$-functions (\ref{beta1}), (\ref{beta2}) gives
\begin{align}
Nx &= (N-8)x^3 - 12x^2y+xy^2 \,, \label{beta1xy} \\
Ny &= -12Nx^3 + 3Nx^2y -9y^3 \,. \label{beta2xy}
\end{align}

These equations have 9 solutions if we allow $x$ and $y$ to be complex.
One of them is the trivial solution $(0,0)$.
Two of them are purely imaginary; they occur at $(0, \pm \sqrt{\frac{N}{9}}i)$.
There are no simple expressions for the remaining six solutions, but it is straightforward to study them numerically.

\begin{figure}[t]
\centering
\includegraphics[width=12cm]{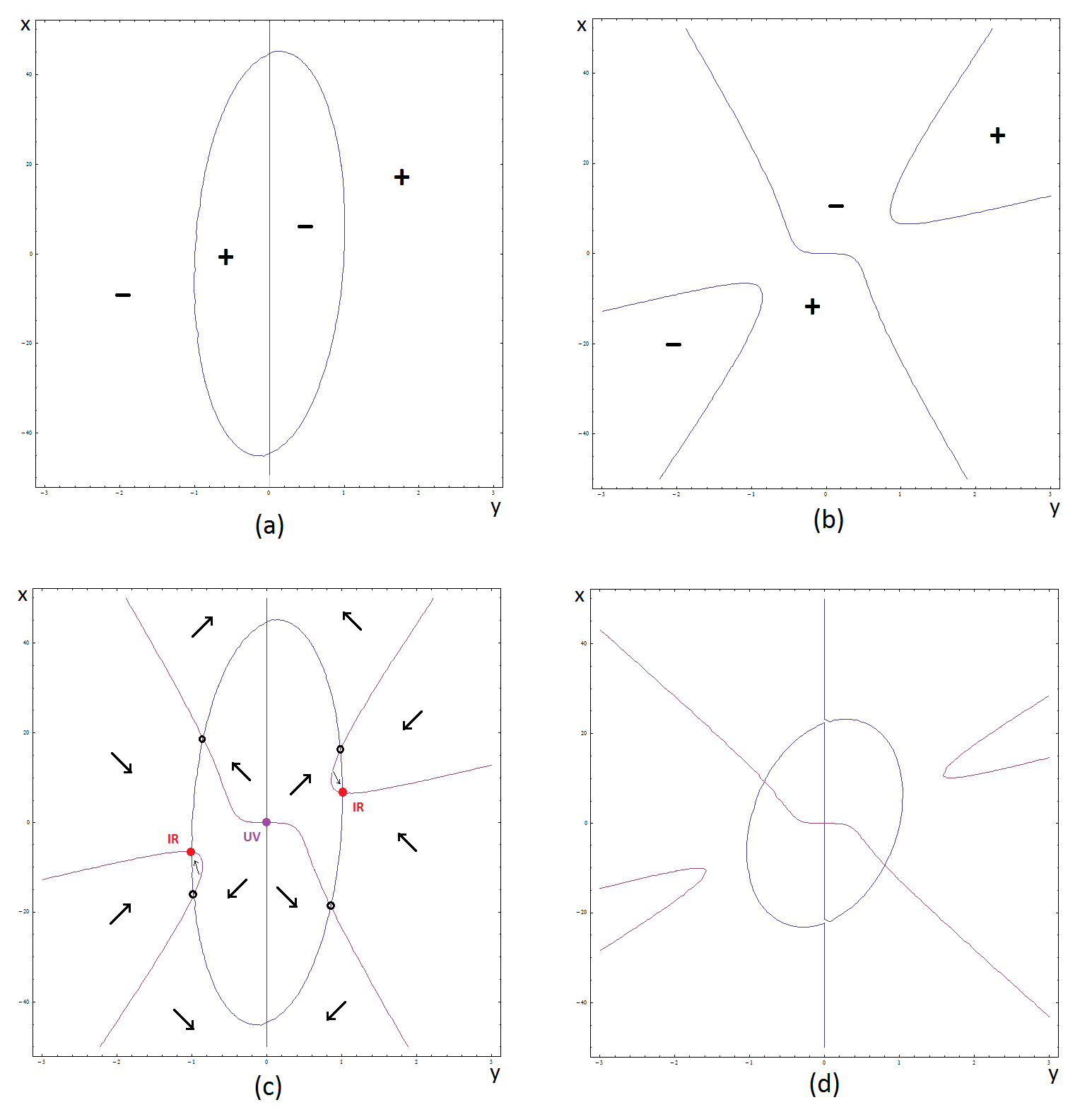}
\caption{(a) zeros of $\beta_1$ at $N=2000$. (b) zeros of $\beta_2$ at $N=2000$. (c) RG flow directions at $N=2000$. (d) only two non-trivial real solutions at $N=500$.}
\label{fig5}
\end{figure}

Two of them are real solutions in the second and fourth quadrant: $(-x_1, y_1)$ and $(x_1, -y_1)$, with $x_1,y_1 > 0$. They exist for any $N$, but they are not
IR stable (they are saddles, i.e. there is one positive and one negative direction in the coupling space). They can be seen in both the bottom left and right graphs in Figure \ref{fig5}, corresponding to $N=2000$ and $N=500$ respectively.

The behavior of the remaining four solutions changes depending on the value of $N$. For $N \leq 1038$, we find that all four solutions are complex.
The bottom right graph of Figure \ref{fig5} shows that for $N=500$ there are indeed only the two real solutions present for any $N$ discussed above.

For $N \geq 1039$, all four of these solutions become real, and they lie in the first and third quadrants: they have the form $(x_3, y_3)$, $(x_4,y_4)$, $(-x_3,-y_3)$, $(-x_4, -y_4)$, with $x_3,y_3,x_4,y_4 >0$. To display the typical behavior of the solutions with $N\geq 1039$, in the bottom left graph of Figure \ref{fig5} we plot the zeroes of $\beta_1$, $\beta_2$ for $N=2000$, with regions of their signs labeled. Combining these, we can get the flow directions in each of these regions in the bottom left graph. We find that for all $N\geq 1039$, we have two stable IR fixed points that are symmetric with respect to the origin, and are labelled as red dots in the figure.
These correspond to the large $N$ solution (\ref{g1star})-(\ref{g2star}), and its equivalent solution with opposite signs of the couplings. For very large $N$, we see that these solutions satisfy $g_2^*=6g_1^*$ as in (\ref{g26g1}). The origin is a UV stable fixed point, and from the direction of the renormalization group flow we can see that all other fixed points are saddle points: they have one stable direction and one unstable direction.

It is interesting to treat $N$ as a continuous parameter and solve for the value $N_{\rm crit}$ where the real IR stable fixed points disappear. First, we notice that we can factor out $x$ in (\ref{beta1xy}), effectively making it quadratic. Moreover, if we subtract (\ref{beta2xy}) from $\frac{y}{x}$ times (\ref{beta1xy}), we will cancel the $Ny$ term, making the second beta function equation a homogeneous equation of order 3. After some more manipulation, we obtain:
\begin{align}
&N = (N-44)x^2+(6x-y)^2\\
&Nx^2(6x-y)-y(4x^2-4y^2+(6x-y)y) = 0 \,.
\end{align}
It is convenient to rewrite these equations in terms of the following variables
\begin{align}
x = \alpha \,,\qquad
y = \beta + 6\alpha \,.
\end{align}
After some algebra, we get:
\begin{align}
&(N-44)\alpha^2+\beta^2 = N\\
&840 \frac{\alpha^3}{\beta^3}+(464-N)\frac{\alpha^2}{\beta^2}+84\frac{\alpha}{\beta}+5 =0 \,.
\end{align}
$N_{\rm crit}$ occurs when the curves defined by the two equations are tangent to each other.
Notice that we have effectively decoupled the two equations, since if we write $z = \alpha/\beta$, we get
\begin{align}
&\alpha^2(N-44+z^{-2}) = N\\
&840z^3 + (464-N)z^2 + 84 z + 5 = 0 \,,
\end{align}
We can just first solve the second equation, then easily solve the first. To determine the critical $N$, we want the second equation to have exactly one real root,
thus we require its discriminant to be zero:
\begin{equation}
\Delta = 18abcd - 4b^3d + b^2c^2 - 4 ac^3 - 27a^2d^2 = 0 \,,
\end{equation}
where $a$, $b$, $c$, $d$ are the coefficients of the cubic equation:
\begin{equation}
a = 840\ ,\quad
b = 464-N \ ,\quad
c = 84 \ ,\quad
d = 5 \,.
\end{equation}
We then arrive at a cubic equation in terms of $N$:
\begin{equation}
\Delta =20N^3-20784N^2+19392N+1856 = 0 \,.
\end{equation}
We can easily write down an analytic expression for $N$, and we find 
\begin{equation}
N_{\rm crit} = 1038.2660492... \,.
\end{equation}
Our numerical solution of the equations is consistent with this value.

For completeness, we now discuss the large $N$ behavior of the four fixed points which are not IR stable (two of them are present for any $N$, and two of them only for $N> N_{\rm crit}$). They are obtained if we assume that, as $N \rightarrow \infty$, $x$ is ${\cal O} (1)$ and $y$ is ${\cal O}(\sqrt{N})$. Then, at the leading order in $N$, we get:
\begin{align}
Nx = Nx^3 + xy^2\,,\qquad
Ny = 3Nx^2y -9y^3 \label{saddle1} \,.
\end{align}
This can be solved to get four solutions: $(\sqrt{\frac{5}{6}}, \sqrt{\frac{1}{6}}\sqrt{N})$, $(-\sqrt{\frac{5}{6}}, \sqrt{\frac{1}{6}}\sqrt{N})$, $(\sqrt{\frac{5}{6}}, -\sqrt{\frac{1}{6}}\sqrt{N})$, $(-\sqrt{\frac{5}{6}}, -\sqrt{\frac{1}{6}}\sqrt{N})$. They correspond to the four real fixed points at large $N$ that are saddle points.
Using the equation from $\beta_1$, we can also check that:
\begin{align}
\Delta_{\sigma} &= 2 - \frac{\eps}{2} + \frac{\eps}{2}\left( \frac{Nx^2+y^2}{N} \right) \nonumber \\
&= 2 \pm \frac{\sqrt{5}}{\sqrt{N}}\eps + {\cal O}(\frac{1}{N}) \,.
\end{align}
The $\frac{\eps}{\sqrt{N}}$ correction does not correspond to the conventional large $N$ behavior.

\section{Operator mixing and anomalous dimensions}
\label{opmix}
\subsection{The mixing of $\sigma^2$ and $\phi^i\phi^i$ operators}
Let us compute the anomalous dimension matrix for operators $\OO^1 = \frac{\phi^i \phi^i}{\sqrt{N}}$ and $\OO^2 = \sigma^2$,
where the $\sqrt{N}$ in the denominator is to ensure that the two point functions of $\OO^1$ and $\OO^2$ are of the same order. They both have classical dimension $4-\eps$ in $d= 6 - \eps$, so we expect them to mix.

\begin{figure}[h]
\centering
\includegraphics{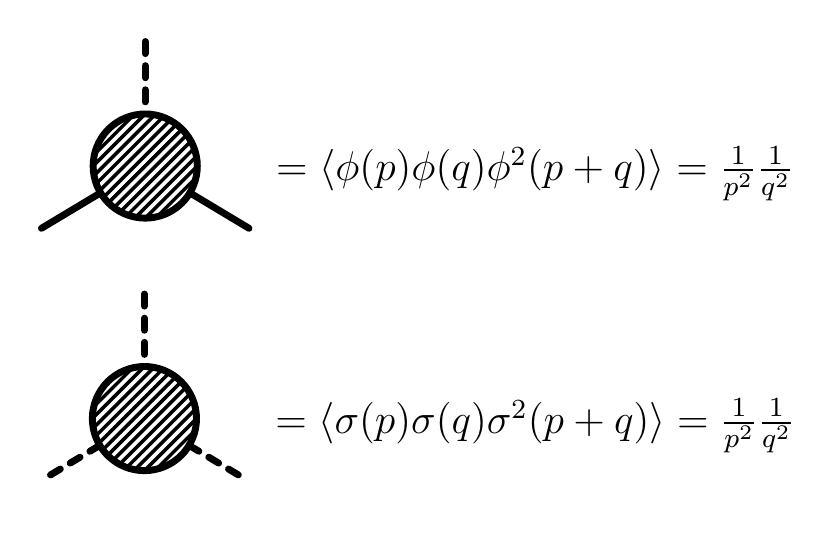}
\caption{Renormalization conditions for the operators $\OO^1$ and $\OO^2$.}
\label{fig6}
\end{figure}

Consider the operators $\OO^1$, and $\OO^2$ renormalized according to the convention shown in Figure \ref{fig6}.

Let $\OO^i_M$ denote the operator renormalized at scale $M$, and $\OO^i_0$ denote the bare operator. We are looking for expressions similar to
(18.53) of \cite{Peskin:1995ev}, i.e. of the form:
\begin{align}
\OO^1_M &= \OO^1_0 + \delta^{11} \OO^1_0+ \delta^{12} \OO^2_0 +  \delta_{\phi}\OO^1_0 \\
\OO^2_M &= \OO^2_0 + \delta^{21} \OO^1_0+ \delta^{22} \OO^2_0 +  \delta_{\sigma}\OO^2_0 \,.
\end{align}

\begin{figure}[h]
\centering
\includegraphics{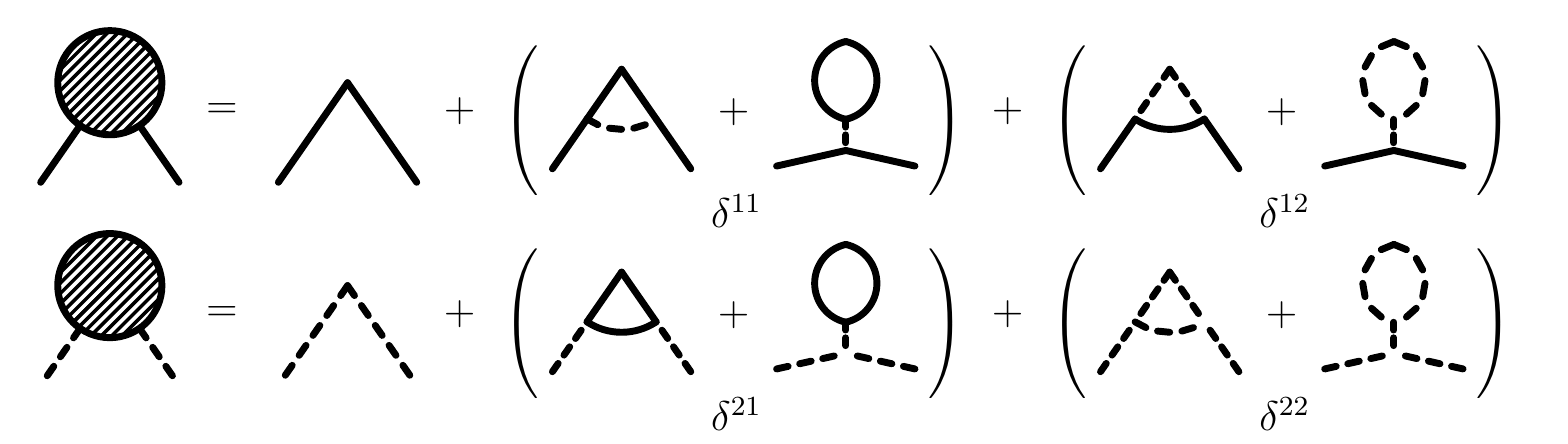}
\caption{Matrix elements of operators $\OO^1$ and $\OO^2$ to 1-loop order.}
\label{fig7}
\end{figure}

The $\delta^{ij}$ counterterms in the above equations are obtained by extracting the logarithmic divergence from the diagrams shown in Figure \ref{fig7}, while the $\delta_{\phi}\OO^1_0$ and  $\delta_{\sigma}\OO^2_0$ terms correspond to external leg corrections, which are omitted in Figure \ref{fig7}.

Each of the terms $\delta^{ij}$ is given by canceling the divergent pieces of two of the above diagrams, as shown.
For example, let us compute the first diagram of the $\delta^{11}$ counter term in Figure \ref{fig7}. We have:
\begin{align}
D1 &= (-g_1)^2\int \frac{d^dk}{(2\pi)^d}\frac{1}{(k-p)^2}\frac{1}{(k+q)^2}\frac{1}{k^2} \nonumber \\
&= (-g_1)^2 I_2 = \frac{g_1^2}{2(4\pi)^{3}}\frac{\Gamma(3-d/2)}{(M^2)^{3-d/2}}\,.
\label{level4D1}
\end{align}
Next, we compute the second diagram of the $\delta^{11}$ counter term. Notice that there is a symmetry factor of 2, and also a factor of $N$ from a closed $\phi$ loop.
\begin{align}
D2 &= \frac{N}{2}(-g_1)^2 \int \frac{d^dk}{(2\pi)^d}\frac{1}{(p+k)^2}\frac{1}{k^2} \frac{1}{p^2}\nonumber \\
&= \frac{Ng_1^2}{2} I_1 \frac{1}{p^2}=-\frac{1}{(4\pi)^{3}}\frac{Ng_1^2}{12} \frac{\Gamma(3-d/2)}{(M^2)^{3-d/2}}\,. \label{level4D2}
\end{align}
These two terms are canceled by $\delta^{11}$; therefore, to leading order in $\epsilon$ we have
\begin{equation}
\delta^{11} = \left( -\frac{g_1^2}{2(4\pi)^3} + \frac{Ng_1^2}{12(4\pi)^3} \right) \frac{\Gamma(3-d/2)}{(M^2)^{3-d/2}}\,.
\end{equation}

Similarly, we can calculate the other three $\delta^{ij}$. The integrals are the same, only the factors of coupling constants and $N$ (due to closed $\phi$ loops) are different. With our normalization convention of $\OO^1$, we need to multiply $\delta^{12}$ by a factor of $\sqrt{N}$, and divide $\delta^{21}$ by a factor of $\sqrt{N}$.
Then the matrix is symmetric, $\delta^{21}= \delta^{12}$, and we find
\begin{equation}
\delta^{12}=\left( -\frac{\sqrt{N}g_1^2}{2(4\pi)^3} + \frac{\sqrt{N}g_1g_2}{1(4\pi)^3}\right) \frac{\Gamma(3-d/2)}{(M^2)^{3-d/2}}\,,
\end{equation}
\begin{equation}
\delta^{22}=\left( -\frac{g_2^2}{2(4\pi)^3} + \frac{g_2^2}{12(4\pi)^3}\right) \frac{\Gamma(3-d/2)}{(M^2)^{3-d/2}}\,.
\end{equation}
Thus, the matrix $\delta^{ij}$ is
\begin{equation}
\delta^{ij} = \frac{1}{12(4\pi)^3} \frac{\Gamma(3-d/2)}{(M^2)^{3-d/2}}\left( \begin{array}{cc}
-6 g_1^2 +Ng_1^2& -6 \sqrt{N} g_1^2 + \sqrt{N}g_1g_2\\
-6\sqrt{N}g_1^2+\sqrt{N}g_1g_2 & -6 g_2^2 + g_2^2\end{array} \right) \,.
\end{equation}
The anomalous dimension matrix is given by
\begin{equation}
\gamma^{ij} = M \frac{\partial}{\partial M}(-\delta^{ij}+\delta_{z}^{ij})\,,
\end{equation}
where we have defined
\begin{equation}
\delta_{z}^{ij} = \left( \begin{array}{cc}
\delta_{\phi} & 0 \\
0 & \delta_{\sigma}  \end{array} \right)\,.
\end{equation}
Now, using expressions for $\delta_{\phi}$ and $\delta_{\sigma}$ from (\ref{deltaz1}), (\ref{deltaz2})\,,
we get:
\begin{equation}
\gamma^{ij} = \frac{-1}{6(4\pi)^3} \left( \begin{array}{cc}
4 g_1^2 -N g_1^2& 6 \sqrt{N} g_1^2 - \sqrt{N}g_1g_2 \\
6 \sqrt{N} g_1^2 - \sqrt{N}g_1g_2 & 4g_2^2-Ng_1^2  \end{array} \right)\,.
\end{equation}
The eigenvalues of this matrix will give the dimensions of the two eigenstates arising from the mixing of operators $\OO^1$ and $\OO^2$.

After plugging in the coupling constants at the IR fixed point from (\ref{g1star}) and (\ref{g2star}), and keeping its entries to order $1/N$, the matrix elements of $\gamma^{ij}$ become:
\begin{equation}
\gamma^{ij} = \frac{-1}{6(4\pi)^3} \left( \begin{array}{cc}
4 g_1^2 -N g_1^2& 6 \sqrt{N} g_1^2 - \sqrt{N}g_1g_2 \\
6 \sqrt{N} g_1^2 - \sqrt{N}g_1g_2 & 4g_2^2-Ng_1^2  \end{array} \right)
\end{equation}
\begin{equation}
=  \eps \left( \begin{array}{cc}
1+\frac{40}{N}& \frac{840}{N^{3/2}} \\
\frac{840}{N^{3/2}} &  1-\frac{100}{N} \end{array} \right) \,.
\end{equation}
To order $1/N$, the off-diagonal terms do not affect the eigenvalues, and we get the scaling dimensions
\begin{equation}
\begin{aligned}
\Delta_{-} = d-2+\gamma_{-} = 4-\frac{100 \epsilon}{N}+{\cal O}\left(\frac{1}{N^2}\right)\\
\Delta_{+}=d-2+\gamma_{+} = 4+\frac{40\epsilon}{N}+{\cal O}\left(\frac{1}{N^2}\right)
\label{delplus}
\end{aligned}
\end{equation}
with the explicit eigenstates given by
\begin{equation}
\OO^+ = \frac{\sqrt{N}}{6} \OO^1 + \OO^2
\end{equation}
\begin{equation}
\OO^- = \frac{-6}{\sqrt{N}} \OO^1 + \OO^2 \,.
\end{equation}
Satisfyingly, we see that to this order the dimension $\Delta_-$ precisely matches the dimension of the only
primary field of dimension near 4 in the large $N$ UV fixed point of the quartic theory. It was determined using large $N$ methods in \cite{Lang:1992zw}, and corresponds to $k=2$ in (\ref{primaries}). Since there are no other primaries of dimension $4+{\cal O}(1/N)$ in the critical theory, we expect that the other dimension $\Delta_{+}$ corresponds to a descendant. Comparing (\ref{delplus}) with (\ref{delsigma}), we see that to this order $\Delta_{+}=2+\Delta_{\sigma}$, so that the eigenstate with eigenvalue $\gamma_{+}$ is indeed a descendant of $\sigma$.

We can actually show that $\Delta_{+}=2+\Delta_{\sigma}$ to all orders in $1/N$ (and for all fixed points) just using the $\beta$-function equations. To start, we redefine our variables as (\ref{g1x}), (\ref{g2y}).
With this definition, the condition satisfied by $x$ and $y$ at the IR fixed point is given in (\ref{beta1xy}), (\ref{beta2xy}), or
\begin{align}
1 &= x^2 - \frac{8}{N}x^2 - \frac{12}{N}xy+\frac{1}{N}y^2 \label{IRfix3} \\
1 &= -12\frac{x^3}{y} + 3x^2 - \frac{9}{N}y^2 \label{IRfix4}\,.
\end{align}
The anomalous dimension matrix in this notation becomes
\begin{equation}
\gamma^{ij}=  \eps \left( \begin{array}{cc}
x^2 - \frac{4}{N}x^2 & \frac{1}{\sqrt{N}}(xy-6x^2) \\
\frac{1}{\sqrt{N}}(xy-6x^2) &  x^2-\frac{4}{N}y^2 \end{array} \right)\,,
\end{equation}
which has eigenvalues
\begin{equation}
\lambda_{\pm} = \frac{(2x^2-\frac{4}{N}x^2-\frac{4}{N}y^2) \pm \sqrt{|\gamma|}}{2}\epsilon,
\end{equation}
where $|\gamma|$ is the determinant, given by:
\begin{align}
|\gamma| = \frac{144}{N}x^4-\frac{48}{N}x^3y + \frac{4}{N}x^2 y^2 + \frac{16}{N^2}x^4 - \frac{32}{N^2}x^2y^2 + \frac{16}{N^2}y^4 \,.
\end{align}
Also, from the first line of (\ref{delsigma}), we see that in terms of $x$ and $y$ the dimension $\Delta_\sigma$ becomes
\begin{equation}
\Delta_\sigma = 2 - \frac{\eps}{2}+\frac{\eps}{2}x^2 + \frac{\eps}{2N}y^2 \,,
\end{equation}
Thus, we have to show that:
\begin{equation}
\Delta_\sigma + 2 = 4 - \eps + \lambda_+
\label{delSigmadesc}
\end{equation}
which, after some algebra, is seen to imply
\begin{equation}
1 - x^2 + \frac{4}{N}x^2 + \frac{5}{N}y^2= \sqrt{|\gamma|}\,.
\end{equation}
Replacing $(1-x^2)$ with (\ref{IRfix3}), we have
\begin{equation}
-\frac{4}{N}x^2 - \frac{12}{N}xy + \frac{6}{N}y^2 = \sqrt{|\gamma|}\,.
\end{equation}
Squaring both sides, and plugging in the expression for $|\gamma|$, we get that the following equation is to hold for (\ref{delSigmadesc}) to be true
\begin{equation}
36x^4 - 12 x^3y + x^2y^2 = \frac{1}{N}\left( 24 x^3y + 32 x^2y^2 - 36 xy^3 + 5 y^4 \right) \,.
\end{equation}
To prove this equality, we again go back to (\ref{IRfix3}) and (\ref{IRfix4}).
If we multiply both of these equations by $3xy-\frac{y^2}{2}$ and subtract the former from the latter, we get exactly the expression above.
Thus, $\Delta_+ = 2 + \Delta_\sigma$ holds to all orders in $1/N$.

\subsection{The mixing of $\sigma^3$ and $\sigma \phi^i\phi^i$ operators}
\label{mixing6}
Next, we would like to calculate the mixed anomalous dimensions of the $\Delta=6$ operators, and show that they are all slightly irrelevant in $d=6-\eps$.
There are six operators with $\Delta = 6$ at $d=6$, they are:
\begin{align}
\OO^1 &= \frac{\sigma (\phi^i)^2}{\sqrt{N}}\ , \qquad
\OO^2 = \sigma^3 \ ,\\
\OO^3 &= \frac{(\partial \phi^i)^2}{\sqrt{N}}\ , \qquad
\OO^4 = (\partial \sigma)^2 \ ,\\
\OO^5 &= \frac{\phi^i \Box \phi^i}{\sqrt{N}}\ , \qquad
\OO^6 = \sigma \Box \sigma.
\end{align}
However, in $d= 6-\eps$, $\OO^1$ and $\OO^2$ have bare dimensions $3\frac{d-2}{2} = 6 - \frac{3}{2}\eps$, while $\OO^3$, $\OO^4$, $\OO^5$, $\OO^6$ have bare dimensions $2+2\frac{d-2}{2} = 6- \eps$. Therefore, in $d= 6-\eps$, $\OO^1$ and $\OO^2$ mix with each other, and $\OO^3$, $\OO^4$, $\OO^5$, $\OO^6$ mix with themselves.

We will compute the mixed anomalous dimensions of $\OO^1$ and $\OO^2$. Using the expressions for the
 $\beta$ functions, (\ref{beta1}), (\ref{beta2}), we can very simply compute the mixed anomalous dimension matrix by differentiating them with respect to
 appropriately normalized couplings (the rescaling is needed because two-point functions of $\OO^1$ and $\OO^2$ are off by a factor of $3N$).
After some algebra, we get:
\begin{equation}
M^{ij}=\left(\begin{array}{cc}
 -\frac{\eps}{2}+\frac{3Ng_1^2 - 24g_1^2-24g_1g_2 + g_2^2}{12(4\pi)^3} & \frac{-12g_1^2+2g_1g_2}{12(4\pi)^3}\sqrt{3N} \\
\frac{-12g_1^2+2g_1g_2}{12(4\pi)^3}\sqrt{3N} & -\frac{\eps}{2}+\frac{Ng_1^2-9g_2^2}{4(4\pi)^3}\end{array} \right) \,.
\label{Mij}
\end{equation}
We plug in the fixed point value of $g_1$ and $g_2$ from (\ref{g1star}) and (\ref{g2star}) to get:
\begin{equation}
M^{ij}=\left( \begin{array}{cc}
\eps - \frac{5040}{N^2}\eps + ... & \frac{840 \sqrt{3}}{N\sqrt{N}} \eps + ...\\
\frac{840 \sqrt{3}}{N\sqrt{N}} \eps + ... & \eps - \frac{420}{N}\eps - \frac{154560}{N^2} \eps + ...\end{array} \right)\,.
\end{equation}
Notice that at this IR fixed point, both eigenvalues of $M^{ij}$ are positive, which implies that the fixed point is IR stable.
This matrix has eigenvalues
\begin{align}
\lambda_1 &= \eps - \frac{3780}{N^2}\eps + {\cal O}(\frac{1}{N^3}) \\
\lambda_2 &= \eps - \frac{420}{N}\eps -\frac{155820}{N^2}\eps + {\cal O}(\frac{1}{N^3})\,.
\end{align}
The relation between scaling dimensions and the eigenvalues of the matrix is given by\footnote{As a simple test of this formula, note that at the free UV fixed point the eigenvalues of (\ref{Mij}) are $\lambda_1=\lambda_2=-\frac{\epsilon}{2}$, giving dimensions $\Delta_1=\Delta_2=3(d/2-1)$ as it should be.}
\begin{equation}
\Delta  = d+ \lambda \,.
\end{equation}
Thus, the scaling dimensions of the mixture of the two operators are:
\begin{align}
\Delta_1 &= d + \lambda_1= 6 + {\cal O} (\frac{1}{N^2})\\
\Delta_2 &= d+ \lambda_2 = 6 - \frac{420}{N} \eps + {\cal O} (\frac{1}{N^2}) \,.
\end{align}
There are no ${\cal O} (\eps)$ correction to the scaling dimensions, as we expected.
As a further check, we note that $\Delta_2$ agrees with the dimension of the $k=3$ primary operator given in (\ref{primaries}).
In the large $N$ Hubbard-Stratonovich approach, this operator is $\sigma^3$ \cite{Lang:1992zw}. Our 1-loop calculation demonstrates the presence of
another primary operator whose dimension is $\Delta_1$. Presumably, the corresponding operator in the Hubbard-Stratonovich approach
is $(\partial_\mu \sigma)^2$.

We can also show that at the other fixed points discussed in Section \ref{finiteN}, $M^{ij}$ has one positive and one negative eigenvalues. If we use (\ref{g1x}), (\ref{g2y}) and plug them into (\ref{Mij}),  we get
\begin{equation}
M^{ij}=\left( \begin{array}{cc}
 -\frac{\eps}{2}+\frac{\eps}{2N}(3Nx^2 - 24x^2 -24 xy + y^2) & \frac{\eps}{2N}(-12x^2+2xy)\sqrt{3N} \\
 \frac{\eps}{2N}(-12x^2+2xy)\sqrt{3N} & -\frac{\eps}{2}+\frac{3\eps}{2N}(Nx^2-9y^2)\end{array} \right) \,.
\label{Mijnew}
\end{equation}
We have shown from solving (\ref{saddle1}) that
\begin{align}
x = \pm \sqrt{5/6}\,,\qquad
y = \pm \sqrt{1/6}\sqrt{N}\,.
\end{align}
Plugging these in, we find that the eigenvalues of $M^{ij}$ at these fixed points are:
\begin{align}
\lambda_1 = \eps\,,\qquad
\lambda_2 = -\frac{5}{3} \eps \ ,
\end{align}
which confirms our graphical analysis that all of these fixed points are saddle points.

\section{Comments on $C_T$ and 5-d $F$ theorem}
\label{cttheorem}
A quantity of interest in a CFT is the coefficient $C_T$ of the stress-tensor 2-point function, which may be defined by
\begin{equation}
\langle T_{\mu\nu}(x_1)T_{\rho\sigma}(x_2)\rangle = C_T \frac{I_{\mu\nu,\rho\sigma}(x_{12})}{x_{12}^{2d}}\,,
\end{equation}
where $I_{\mu\nu,\rho\sigma}(x_{12})$ is a tensor structure uniquely fixed by conformal symmetry, see e.g. \cite{Osborn:1993cr}. For $N$ free real scalar fields in dimension $d$ with canonical normalization, one has $C_T = \frac {Nd}{(d-1)S_d^2}$, where $S_d$ is the volume of the $d$-dimensional round sphere. In the critical $O(N)$ theory, using large $N$ methods one finds the result \cite{Petkou:1994ad, Petkou:1995vu}
\begin{equation}
\begin{aligned}
&C_T = \frac{Nd}{(d-1)S_d^2}\left(1+\frac{1}{N}C_{T,1}+{\cal O}(\frac{1}{N^2})\right)\\
&C_{T,1}= -\left(\frac{2{\cal C}(\mu)}{\mu+1}+\frac{\mu^2+3\mu-2}{\mu(\mu^2-1)}\right)\eta_1\,,\qquad \mu=\frac{d}{2} \label{CTcorr}
\end{aligned}
\end{equation}
where ${\cal C}(\mu) = \psi(3-\mu)+\psi(2\mu-1)-\psi(1)-\psi(\mu)$, and $\psi(x)=\frac{\Gamma'(x)}{\Gamma(x)}$.
It is interesting to analyze the behavior of $C_{T,1}$ in $d=6-\epsilon$. The anomalous dimension $\eta_1$, given in (\ref{eta1-phi}), is of order $\epsilon$, but there is a pole in ${\cal C}(\mu)$ from the term $\psi(3-\mu) \sim -2/\epsilon$. Hence, one finds
at $d=6-\epsilon$
\begin{equation}
C_{T,1} = 1-\frac{7}{4}\epsilon+{\cal O}(\epsilon^2)
\end{equation}
and so
\begin{equation}
C_T = \frac{d}{(d-1)S_d^2}\left(N+1-\frac{7}{4}\epsilon\right)+{\cal O}(\epsilon^2)\,.
\label{CT-6d}
\end{equation}
Thus, as $d\rightarrow 6$, we find the $C_T$ coefficient of $N+1$ free real scalars. This provides a nice check on our description of the critical $O(N)$
theory via the IR fixed point of (\ref{cubic6d}). The ${\cal O}(\epsilon)$ correction may be calculated in this description as well,
using the 1-loop fixed point (\ref{g1star}), (\ref{g2star}), but we leave this for future work.

\begin{figure}
\begin{center}
\includegraphics[width=8cm]{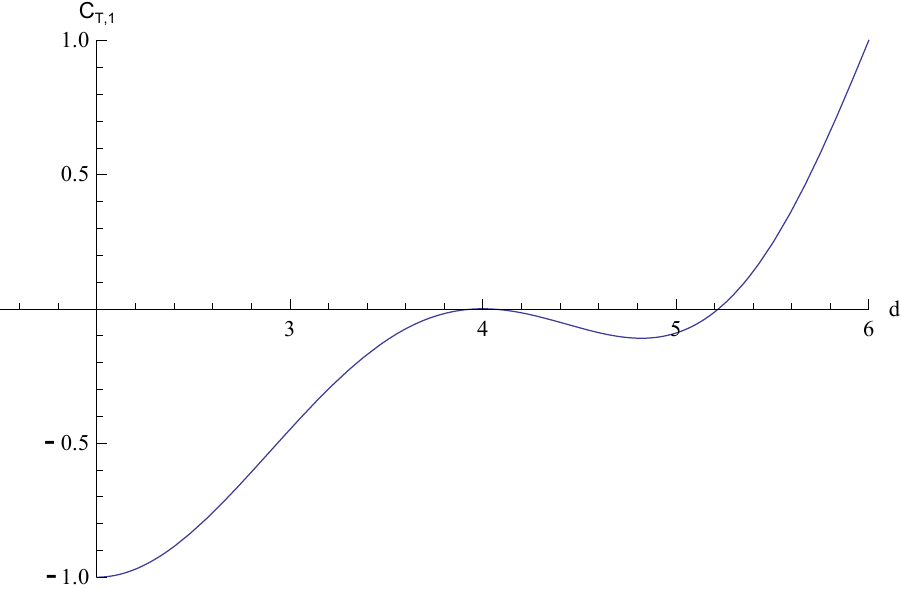}
\caption{The ${\cal O} (N^0)$ correction $C_{T,1}$ in the coefficient
$C_T$ of the stress tensor two-point function in the critical $O(N)$ theory for $2<d<6$; see (\ref{CTcorr}).
Note that $C_{T,1}$ is negative for $2<d \lesssim 5.22$, so that in this range of dimensions $C_{T}^{\rm crit} < N C_T^{\rm free~sc.}$ for large $N$. Therefore, the $C_T$ theorem is respected in $2<d<4$, but violated for $4<d\lesssim 5.22$.}
\label{CT1}
\end{center}
\end{figure}

 The appearance of an extra massless scalar field as $d\rightarrow 6$ is also suggested by dimensional analysis. The dimension of $\sigma$ is $2+{\cal O}(1/N)$ in all $d$, so as $d$ approaches $6$, it becomes the dimension of a free scalar field. Then the two-derivative kinetic term for $\sigma$ (as well as the $\sigma^3$ coupling) become classically marginal. Note also that the negative sign of the order $\epsilon$ correction in (\ref{CT-6d}) implies that $C_T$ decreases from the UV fixed point of $N+1$ free fields to the interacting IR fixed point, consistently with the idea that $C_T$ may be a measure of degrees of freedom \cite{Petkou:1995vu}. However, expanding $C_T$ in $d=4+\epsilon$, one finds
\begin{equation}
C_T = \frac{d}{(d-1)S_d^2}\left(N-\frac{5}{12}\epsilon^2\right)+{\cal O}(\epsilon^3)\,.
\label{CT-4d}
\end{equation}
According to our interpretation, the critical theory in $d=4+\epsilon$ should be indentified with the UV fixed point of the $-\phi^4$ theory, while the IR fixed point corresponds to $N$ free scalars. Then, (\ref{CT-4d}) is seen to violate $C_T^{UV}>C_T^{IR}$. A plot of $C_{T,1}$ in the range $2<d<6$ is given in Figure \ref{CT1}. Finally, let us quote the value of $C_T$ in the interesting dimension $d=5$
\begin{equation}
C_T^{d=5}=\frac{5}{4 S_5^2}\left(N-\frac{1408}{1575\pi^2}\right)\,.
\end{equation}
Again, we observe that this value is consistent with $C_T^{UV}>C_T^{IR}$ for the flow from the free UV theory of $N+1$ massless scalars to the interacting fixed point, but not for the flow from the interacting fixed point to the free IR theory of $N$ massless scalars. Thus, the latter flow provides a non-supersymmetric counterexample against the possibility of a $C_T$ theorem (for a supersymmetric counterexample, see \cite{Nishioka:2013gza}).
In contrast, the 5-dimensional F-theorem holds for both flows, as we discuss below.

It was proposed in \cite{Klebanov:2011gs} (see also \cite{Myers:2010tj,Casini:2011kv}) that, for any odd dimensional Euclidean CFT, the quantity
\begin{equation}
\tilde F = (-1)^{\frac{d+1}{2}} F = (-1)^{\frac{d-1}{2}} \log Z_{S^d}\,,
\end{equation}
should decrease under RG flows
\begin{equation}
\tilde F_{UV} > \tilde F_{IR}\,.
\label{tildeF}
\end{equation}
Here $F=-\log Z_{S^d}$ is the free energy of the CFT on the round sphere, which is a finite number after the power law UV divergences are regulated away (for instance, by $\zeta$-function or dimensional regularization). Note that when we put the CFT on the sphere, we have to add the conformal coupling of the scalar fields to the $S^d$ curvature. This effectively renders the vacuum metastable, both in the case of the $-\phi^4$ theory and in the cubic theory. Similarly, the CFT is metastable on $R\times S^{d-1}$, which is relevant for calculating the scaling dimensions of operators.

For $d=3$ (where $\tilde F = F$), the F-theorem (\ref{tildeF}) was proved in \cite{Casini:2012ei}. However, higher dimensional versions
are less well established. Using the results of this paper, we can provide a simple new test of the 5d version of the F-theorem (in this case $\tilde F=-F$). As we have argued, the 5d critical $O(N)$ theory can be viewed as either the IR fixed point of the cubic theory (\ref{cubic6d}) or the UV fixed point of the quartic scalar theory. This implies that $\tilde F$ should satisfy
\begin{equation}
N \tilde F_{\rm free~sc.} < \tilde F_{\rm crit.} < (N+1)\tilde F_{\rm free~sc.}\,,
\label{5dF}
\end{equation}
where $\tilde F_{\rm free~sc.}$ is minus the free energy of a 5d free conformal scalar \cite{Klebanov:2011gs}
\begin{equation}
\tilde F_{\rm free~sc.} = \frac{\log 2}{128}+\frac{\zeta(3)}{128\pi^2}-\frac{15\zeta(5)}{256\pi^4}\simeq 0.00574\,.
\label{5dscalar}
\end{equation}
The value of $F$ at the critical point can be calculated perturbatively in $1/N$ by introducing the Hubbard-Stratonovich auxiliary field as reviewed in Section 2. The leading ${\cal O} (N)$ term is the same as in the free theory, while the ${\cal O} (N^0)$ term arises from the determinant of the non-local kinetic operator for the auxiliary field \cite{Gubser:2002vv, Diaz:2007an, Klebanov:2011gs}. The result is \cite{Giombi:2014iua}
\begin{equation}
\tilde F_{\rm crit.} = N \tilde F_{\rm free~sc.} + \frac{3\zeta(5)+\pi^2\zeta(3)}{96\pi^2}+{\cal O}\left(\frac{1}{N}\right)\,.
\label{5dcrit}
\end{equation}
The same answer may be obtained by computing the ratio of determinants for the bulk scalar field in the $AdS_6$ Vasiliev theory with alternate boundary conditions \cite{Giombi:2014iua}. We see that the left inequality in (\ref{5dF}) is satisfied, since the ${\cal O} (N^0)$ correction in (\ref{5dcrit}) is positive (this check was already made in \cite{Giombi:2014iua}). More non-trivially, we observe that the right inequality also holds, because
$\frac{3\zeta(5)+\pi^2\zeta(3)}{96\pi^2} \simeq 0.001601$ is smaller than the value of $\tilde F$ for one free scalar, eq. (\ref{5dscalar}).

\section{Gross-Neveu-Yukawa model and a test of 3-d F-theorem}
\label{GNYukawa}

The action of the Gross-Neveu (GN) model \cite{Gross:1974jv} is given by
\begin{equation}
S(\bar{\psi},\psi) = -\int d^d x\left[\bar{\psi}^i \slashed{\partial} \psi^i + \frac{1}{2}g(\bar{\psi}^i \psi^i)^2  \right]\,.
\end{equation}
Here $N=\tilde N \rm tr \bf 1$, where $\rm tr \bf 1$ is the trace of the identity in the Dirac matrix space, and $\tilde N$ is the number of Dirac fermion fields $\psi^i$ ($i=1,\ldots,\tilde N$). The parameter $N$ counts the actual number of fermion components, and it is the natural parameter to write down the $1/N$ expansion (factors of $\rm tr \bf 1$ never appear in the expansion coefficients if $N$ is used as the expansion parameter).

The beta functions and anomalous dimensions of this model in $d = 2+ \eps$ can be calculated to be (see, for instance, \cite{Moshe:2003xn})
\begin{align}
\beta(g) &= \eps g - (N-2) \frac{g^2}{2\pi} + (N-2)\frac{g^3}{4\pi^2}+{\cal O}(g^4) \\
\eta_{\psi}(g) &= \frac{N-1}{8\pi^2}g^2 - \frac{(N-1)(N-2)}{32\pi^3}g^3+ {\cal O}(g^4) \\
\eta_{\cal M} (g) &= \frac{N-1}{2\pi} g -\frac{N-1}{8\pi^2}g^2 + {\cal O}(g^3),
\end{align}
where $\eta_{\cal M}$ is related to the anomalous dimension of the composite field $\sigma = \bar{\psi}\psi$. We can solve the beta function for the critical value of $g$ at the fixed point:
\begin{equation}
g^* = \frac{2\pi}{N-2}\eps \left( 1 - \frac{\eps}{N-2}\right) + {\cal O}(\eps^3)\ .
\end{equation}
Plugging this value of $g^*$, we can find the dimensions of the fermion field and the composite field:
\begin{align}
\Delta_{\psi} &= d-1 + \eta_{\psi}(g^*) = \frac{1 + \eps}{2} + \frac{N-1}{4(N-2)^2}\eps^2 + {\cal O}(\eps^3) \ ,\\
\Delta_{\sigma} &= d-1 - \eta_{\cal M}(g^*) = 1 - \frac{\eps}{N-2}+{\cal O}(\eps^2)\ .
\label{anom-GN}
\end{align}

The UV fixed point of the GN model in $2<d<4$ dimensions is related to the IR fixed point of the
Gross-Neveu-Yukawa (GNY) model, which has the following action
\cite{Hasenfratz:1991it,ZinnJustin:1991yn,Moshe:2003xn}
\begin{equation}
S(\bar{\psi},\psi,\sigma) = \int d^d x \left[ -\bar{\psi}^i  \left( \slashed{\partial} + g_1 \sigma \right) \psi^i + \frac{1}{2} (\partial_{\mu} \sigma)^2 + \frac{g_2}{24}\sigma^4 \right].
\end{equation}
In $d=4-\eps$, the one-loop beta functions of the GNY model are given by:
\begin{align}
\beta_{g_1^2} &= -\eps g_1^2 + \frac{N+6}{16\pi^2}g_1^4 \\
\beta_{g_2} &= -\eps g_2 + \frac{1}{8\pi^2} \left(\frac{3}{2} g_2^2 + N g_2 g_1^2 - 6 N g_1^4 \right) \ .
\end{align}
For small $\eps$, one finds that there is a stable IR fixed point for any positive $N$:
\begin{align}
(g_{1}^*)^2 &= \frac{16 \pi^2 \eps}{N+6} \\
g_{2}^* &= 16 \pi^2 R \eps,
\end{align}
where
\begin{equation}
R = \frac{24N}{(N+6)[(N-6)+\sqrt{N^2 + 132N +36}]}.
\end{equation}
The anomalous dimensions of the elementary fields are given by
\begin{align}
\gamma_{\psi} = \frac{1}{32\pi^2}g_1^2 \,,\qquad
\gamma_{\sigma} = \frac{N}{32\pi^2}g_1^2.
\end{align}
and plugging in the value of the fixed point couplings, one obtains
\begin{align}
\Delta_{\psi} &= \frac{d-1}{2} + \gamma_{\psi}
= \frac{3}{2} - \frac{N+5}{2(N+6)}\eps \\
\Delta_{\sigma} & = \frac{d-2}{2} + \gamma_{\sigma}
= 1 - \frac{3}{N+6} \eps\,.
\label{anom-GNY}
\end{align}

The large $N$ expansion of the 3d critical fermion theory may be developed in arbitrary dimension following similar lines as in the scalar case reviewed in Section 2: one introduces an auxiliary field $\sigma$ to simplify the quartic interaction, and develops the $1/N$ perturbation theory with the effective non-local propagator for $\sigma$ and the $\sigma\bar\psi \psi$ interaction. The anomalous dimensions of $\psi$ and $\sigma$ in the critical theory have been calculated respectively to order $1/N^3$ and $1/N^2$, and in arbitrary dimension $d$ \cite{Derkachov:1993uw, Gracey:1992cp}. To leading order in $1/N$, the explicit expressions are given by
\begin{eqnarray}
\Delta_{\psi} &=& \frac{d-1}{2}-\frac{\Gamma\left(d-1\right)\sin(\frac{\pi d}{2})}{\pi\Gamma\left(\frac{d}{2}-1\right)\Gamma\left(\frac{d}{2}+1\right)}\frac{1}{N}+{\cal O}\left(\frac{1}{N^2}\right)\\
\Delta_{\sigma} &=& 1+\frac{4(d-1)\Gamma\left(d-1\right)\sin(\frac{\pi d}{2})}{\pi(d-2)\Gamma\left(\frac{d}{2}-1\right)\Gamma\left(\frac{d}{2}+1\right)}\frac{1}{N}+{\cal O}\left(\frac{1}{N^2}\right)\,.
\end{eqnarray}
Expanding these expressions in $d=2+\epsilon$ and $d=4-\epsilon$, one can verify the agreement with (\ref{anom-GN}) and (\ref{anom-GNY}) respectively. Note that both
$\Delta_{\psi}$ and $ \Delta_{\sigma}$
go below their respective unitarity bounds for $d>4$; so, unlike in the scalar case, there is no unitary critical fermion theory above dimension four.
One may check that the available subleading large $N$ results \cite{Derkachov:1993uw, Gracey:1992cp} also precisely agree with the $1/N$ expansion of (\ref{anom-GN}) and (\ref{anom-GNY}), giving strong support to the fact that the 3d critical fermionic CFT may be viewed as either the IR fixed point of the GNY model, or the UV fixed point of the GN theory.

We may use the two alternative descriptions of the critical fermion theory to provide a simple test of the 3d F-theorem, similar to the tests carried out in \cite{Klebanov:2011gs}. In the UV, the GNY model (which has relevant interactions in $d=3$) is a free theory of $N$ fermions and one conformal scalar, while the GN model is free in the IR. Thus, the value of $F$ for the critical fermion theory should satisfy
\begin{equation}
N F_{\rm free~fermi} < F_{\rm crit.~fermi} < N F_{\rm free~fermi}+F_{\rm free~sc.}\,.
\label{3dF}
\end{equation}
In $d=3$, one finds
\begin{eqnarray}
F_{\rm free~sc.} = \frac{\log 2}{8}-\frac{3\zeta(3)}{16\pi^2}\simeq 0.0638\,.
\end{eqnarray}
The value of $F$ in the critical theory may be computed perturbatively in the $1/N$ expansion, and one obtains the result \cite{Diaz:2007an,Klebanov:2011gs}
\begin{equation}
F_{\rm crit.~fermi} =  N F_{\rm free~fermi}+\frac{\zeta(3)}{8\pi^2}+{\cal O}\left(\frac{1}{N}\right)\,.
\end{equation}
Because $\frac{\zeta(3)}{8\pi^2} \simeq 0.0152 < F_{\rm free~sc.}$, we see that indeed (\ref{3dF}) is satisfied in both directions.

\section*{Acknowledgments}

We thank David Gross, Daniel Harlow, Igor Herbut, Juan Maldacena, Giorgio Parisi, Silviu Pufu, Leonardo Rastelli, Dam Son, Grigory Tarnopolsky and Edward Witten for useful discussions.
The work of LF and SG was supported in part by the US NSF under Grant No.~PHY-1318681. The work of IRK was supported in part by the US NSF under Grant No.~PHY-1314198.

\appendix

\section{Some useful integrals}
Let us evaluate the following  useful one-loop integral in dimensional regularization
\begin{eqnarray}
I(\alpha,\beta) &=& \int \frac{d^dq}{(2\pi)^d} \frac{1}{q^{2\alpha} (p+q)^{2\beta}} \cr
&=&\int_0^1 dx x^{\alpha-1}(1-x)^{\beta-1}\frac{\Gamma\left(\alpha+\beta\right)}{\Gamma(\alpha)\Gamma(\beta)} \int \frac{d^dq}{(2\pi)^d} \frac{1}{\left[q^2+x(1-x)p^2\right]^{\alpha+\beta}}\cr
&=& \int_0^1 dx \frac{x^{\alpha-1}(1-x)^{\beta-1}}{\Gamma(\alpha)\Gamma(\beta)}
\int_0^{\infty} dt t^{\alpha+\beta-1}\,\int \frac{d^dq}{(2\pi)^d} e^{-t\left(q^2+x(1-x)p^2\right)}\cr
&=&\frac{1}{(4\pi)^{\frac{d}{2}}}\frac{\Gamma\left(\frac{d}{2}-\alpha\right)\Gamma\left(\frac{d}{2}-\beta\right)\Gamma\left(\alpha+\beta-\frac{d}{2}\right)}
{\Gamma\left(\alpha\right)\Gamma\left(\beta\right)\Gamma\left(d-\alpha-\beta\right)}\left(\frac{1}{p^2}\right)^{\alpha+\beta-\frac{d}{2}}
\label{Iab}
\end{eqnarray}
In the calculations in $d=6-\epsilon$, we often encounter the case $\alpha=1,\beta=1$, which gives
\begin{equation}
I_1 \equiv I(1,1) = \frac{1}{(4\pi)^{\frac{d}{2}}} \frac{\Gamma\left(3-\frac{d}{2}\right)\Gamma\left(\frac{d}{2}-1\right)^2}{(2-\frac{d}{2})\Gamma\left(d-2\right)}\left(\frac{1}{p^2}\right)^{2-\frac{d}{2}}
\end{equation}
For the purpose of extracting the logarithmic terms in $d=6-\epsilon$, it is sufficient to use the approximation
\begin{equation}
I_1 \rightarrow  -\frac{p^2}{6(4\pi)^{3}}\frac{\Gamma(3-d/2)}{(p^2)^{3-d/2}}\ .
\label{I1}
\end{equation}

We will also need the following three-propagator integral
\begin{align}
I_2 &= \int \frac{d^d k}{(2\pi)^d}\frac{1}{(k-p)^2}\frac{1}{(k+q)^2}\frac{1}{k^2} \\
&= 2 \int_0^1 dxdydz \delta(1-x-y-z) \int \frac{d^d k}{(2\pi)^d} \frac{1}{[x(k-p)^2+y(k+q)^2+zk^2]^3} \\
&= 2 \int_0^1 dxdydz \delta(1-x-y-z) \int \frac{d^d k}{(2\pi)^d} \frac{1}{[k^2 - 2xkp + 2ykq + xp^2+yq^2]^3}
\end{align}
Defining $l \equiv k -xp + yq$, we get:
\begin{equation}
I_2 =  2 \int_0^1 dxdydz \delta(1-x-y-z) \int \frac{d^d l}{(2\pi)^d} \frac{1}{[l^2 + \Delta]^3}
\end{equation}
with $\Delta = x(1-x)p^2+y(1-y)q^2+ 2xy pq$.
Evaluating the standard momentum integral, we obtain
\begin{align}
I_2 &= \int_0^1 dxdydz \delta(1-x-y-z) \frac{1}{(4\pi)^{d/2}}\frac{\Gamma(3-d/2)}{\Delta^{3-d/2}}\,.
\end{align}
In the RG calculations of Section 3, we use the renormalization conditions \cite{Peskin:1995ev} $p^2 = q^2 = (p+q)^2 = M^2$, which imply
$2p\cdot q = -M^2$, and hence $\Delta = M^2(x(1-x)+y(1-y)-xy)$. So we can write
\begin{equation}
I_2 = \frac{1}{(4\pi)^{d/2}}\frac{\Gamma(3-d/2)}{(M^2)^{3-d/2}}\int_0^1 dxdydz \delta(1-x-y-z)\frac{1}{\left(x(1-x)+y(1-y)-xy\right)^{3-d/2}}\,.
\end{equation}
In $d=6-\epsilon$
the Feynman parameter integral becomes
\begin{equation}
\int_0^1 dxdydz \delta(1-x-y-z)\left ( 1 - \frac \epsilon 2 \log (x(1-x)+y(1-y)-xy) + {\cal O}(\epsilon^2) \right ) = \frac{1}{2} + {\cal O} (\epsilon)\ .
\end{equation}
Thus,
\begin{equation}
I_2=
\frac{1}{2(4\pi)^{3}} \left ( \frac{2}{\epsilon}- \log M^2 + A + {\cal O} (\epsilon) \right ) \ ,
\label{I2first}
\end{equation}
where $A$ is an unimportant constant. For the purpose of extracting the $\log M^2$ terms in $d=6-\epsilon$, it is sufficient to use the approximation
\begin{equation}
I_2 \rightarrow
\frac{1}{2(4\pi)^{3}}\frac{\Gamma(3-d/2)}{(M^2)^{3-d/2}}\ .
\label{I2}
\end{equation}


\bibliographystyle{ssg}
\bibliography{CGLP}

\end{document}